\documentclass{aa}
\bibpunct{(}{)}{;}{a}{}{,}
\usepackage[varg]{txfonts}
\usepackage{graphicx,gensymb}
\usepackage{subcaption}
\usepackage[normalem]{ulem}
\usepackage{amsmath, amssymb, amsfonts}
\usepackage{rotating} 
\usepackage{adjustbox}
\usepackage{ulem}
\usepackage{upgreek}
\usepackage{float}
\begin{document}
\pagestyle{empty}

\title{An empirical model of the extragalactic radio background}

\author{Fangyou Gao\inst{\ref{inst1},\ref{inst2}}
\and Tao Wang\thanks{\email{taowang@nju.edu.cn}}\inst{\ref{inst1},\ref{inst2}}
\and Yijun Wang\inst{\ref{inst1},\ref{inst2}}
}
\institute{School of Astronomy and Space Science, Nanjing University, 163 Xianlin Avenue, Nanjing 210023, People's Republic of China\label{inst1}
\and 
Key Laboratory of Modern Astronomy and Astrophysics (Nanjing University), Ministry of Education, Nanjing 210023, People's Republic of China\label{inst2}
}

\abstract{}{Radio observations provide a powerful tool to constrain the assembly of galaxies over cosmic time. Recent deep and wide radio continuum surveys have improved significantly our understanding on radio emission properties of active galactic nuclei (AGNs) and star-forming galaxies (SFGs) across $0 < z < 4$. This allows us to 
derive an empirical model of the radio continuum emission of galaxies based on their star formation rates and the probability of hosting an radio AGN. 
Here we verify how well this empirical model can reproduce the extragalactic radio background (ERB), which can provide new insights into the contribution to the ERB from galaxies of different masses and redshfits. }
{We make use of the Empirical Galaxy Generator (EGG) to generate a near-infrared (NIR)-selected, flux-limited multi-wavelength catalog to mimic real observations. Then we assign radio continuum flux densities to galaxies based on their star formation rates and the probability of hosting a radio-AGN of specific 1.4 GHz luminosity. We also apply special treatments to reproduce the clustering signal of radio AGNs.}
{Our empirical model successfully recovers the observed 1.4 GHz radio luminosity functions (RLFs) of both AGN and SFG populations, as well as the differential number counts at various radio bands. The uniqueness of this approach also allows us to directly link radio flux densities of galaxies to other properties, including redshifts, stellar masses, and magnitudes at various photometric bands. We find that roughly half of the radio continuum sources to be detected by the Square Kilometer Array (SKA) at $z \sim 4-6$ will be too faint to be detected in the optical survey ($r \sim 27.5$) carried out by Rubin observatory.}
{Unlike previous studies which utilized (extrapolations of) RLFs to reproduce ERB, our work starts from a simulated galaxy catalog with realistic physical properties. It has the potential to simultaneously, and self-consistently reproduce physical properties of galaxies across a wide range of wavelengths, from optical, NIR, far-infrared (FIR) to radio wavelengths. Our empirical model can shed light on the contribution of different galaxies to the extragalactic background light, and greatly facilitates designing future multiwavelength galaxy surveys.}

\authorrunning{F.Y. Gao et al.}
\titlerunning{RULES I: an empirical model of the extragalactic radio background}
\maketitle

\section{Introduction}\label{intro}
Being one of the spectral windows that are less affected by Earth's atmosphere, radio observations are advantageous in operating at a wide range of frequencies and probing a wide range of redshifts. Hence, radio observations serve as a powerful tool to constrain the growth and evolution of galaxies across the history of our universe. Our knowledge regarding the extragalactic radio background has accumulated thanks to many generations of radio surveys. It has been established that powerful radio sources are usually radio emission produced by accreting supermassive black boles, which are expected to be responsible for the regulation to their host galaxies \citep[e.g.,][]{2006MNRAS.365...11C}, while less powerful radio sources are predominately star forming galaxies (SFGs), displaying a tight correlation between radio emission and star formation rates (SFRs). This knowledge is applied in many fields, for example, selecting radio active galactic nuclei (AGNs) according to radio luminosity \citep[see][for a recent review]{2016A&ARv..24...10T} and using radio emission as a dust-invariant indicator of SFRs \citep[e.g.,][among others]{2001ApJ...554..803Y,2009MNRAS.397.1101G,2017ApJ...847..136B,2017MNRAS.466.2312D,2019A&A...631A.109W}.

Realistic simulations of extragalactic radio galaxies and images provide valuable support in many aspects, such as evaluating the performance of both "hardware" instruments and "software" source extraction pipelines. Popular mock radio source catalogs are mainly built upon (extrapolations of) radio luminosity functions (RLFs). As a straightforward measurement, RLF reflects the abundance of radio sources within a certain luminosity range across a wide range of redshift. Therefore, it is convenient to directly calculate how many radio sources should exist within a certain flux density range at any given redshift. 

Through this method, current mock radio catalogs have succeeded in reproducing extragalactic radio background (ERB) based on RLFs of different populations. For example, SKA Design Study Simulated Skies simulation \citep[$S^{3}$;][]{2008MNRAS.388.1335W} produced mock radio sources based on observed RLFs of five distinct galaxy populations: radio-quiet AGNs, radio-loud AGNs composed of FRI and FRII sources, and star forming galaxies (SFGs) composed of quiescent and star-bursting galaxies. More recent work by The Tiered Radio Extragalactic Continuum Simulation \citep[TRECS;][]{2019MNRAS.482....2B} exploited RLF models developed in \citet{2010MNRAS.404..532M} and split radio-loud AGNs into three populations with different evolutionary properties: steep-spectrum sources (SS-AGNs), flat-spectrum radio quasars (FSRQs) and BL Lacs. These mock radio catalogs prove to be successfully recover many observable quantities such as RLFs and source number counts at various frequencies. However, it is difficult to gain insights into other physical information of these radio sources, such as how quickly they form stars and how they shine at other wavelenghths. This empirical method also overlooked physical origin behind the scene, for example, how the fraction of radio AGNs evolve across cosmic time as well as how much contribution to ERB made by radio AGNs and SFGs separately at given stellar mass.

Considering this issue, we initiate this work, aiming to recover ERB not based on observed RLFs but based on mock mass-complete galaxy catalog, which starts from fully-explored stellar mass functions and incorporate rich information on physical properties.
 
Since the past few decades, we have explored deeper and deeper in our universe and hence established comprehensive knowledge towards the physical properties of galaxies, at least at $0<z<4$, such as the redshift-dependent distribution of their stellar masses\citep[e.g.,][among many others]{2013ApJ...777...18M,2014MNRAS.444.2960D,2014ApJ...783...85T,2017A&A...605A..70D} and star formation rates \citep[SFRs, e.g.,][among many others]{2007ApJ...660L..43N,2014ApJS..214...15S,2015A&A...575A..74S,2018A&A...615A.146P}, as well as spectral energy distributions (SEDs) of different populations across a wide range of spectrum \citep[e.g.,][among many others]{2002ApJ...576..159D,2012ApJ...760....6M,2013ARA&A..51..393C}. By utilizing these empirical knowledge, astronomers are able to generate mock galaxy catalogs, reproduce various observables in a self-consistent way and extrapolate to much deeper depth,  predicting what type of galaxies can be directly observed (e.g., SIDES simulation of far-infrared to submillimeter extragalactic background in \citet{2017A&A...607A..89B} and EGG code \citet{2017A&A...602A..96S} used in this work). This method is also capable of creating mock galaxy catalogs at a fast speed, without inducing complex large-scale dark matter simulations which are computationally expensive \citep{2017A&A...602A..96S}. 

Through this method, we are able to connect radio sources to their host galaxies with a plethora of physical information, enabling studying extragalactic radio sources in a multi-wavelength context and harnessing the power of combining different observational techniques.



%

The next-generation radio facility, the Square Kilometer Array (SKA) is expected to revolutionize our current knowledge regarding finding fainter and farther radio sources with its supreme sensitivity and angular resolution. Before its final operation, it is necessary to assess how far it can surpass the current radio telescopes and what improvement it can bring when work in collaboration with other mainstream optical\textbackslash{}near-infrared (NIR) large sky surveys such as The Large Synoptic Survey Telescope \citep[LSST,][]{2019ApJ...873..111I}. Therefore, it is of vital importance to build new models to predict what type of galaxies that will be detected with different observation time, help to design observational strategies for different scientific goals and organize surveys at other bands in a synergistic way. Our mock catalog, being successfully recover both radio and multiwavelengh sky self-consistently, will provide essential help to this subject.

The structure of this letter is as follows: we describe our simulation setups in Section \ref{setup}, and validate our results in Section \ref{validation}. In Section \ref{app} we demonstrate the various applications of our simulations and discuss some deficiencies in Section \ref{def}. We briefly summarize our work in Section \ref{summary}. Throughout this letter, we assume a flat $\Lambda$CDM universe with $\Omega_{\text{M}} = 0.286$ and $H_{0}=69.3 \,\rm\ km\ s^{-1}\ Mpc^{-1}$ \citep[Nine-year Wilkinson Microwave Anisotropy Probe (WMAP) results;][]{2013ApJS..208...19H}. Unless otherwise stated, we adopt a \citet{1955ApJ...121..161S} initial mass function (IMF). For magnitudes, we use standard AB magnitude system.

\section{Simulation setups}\label{setup}
\subsection{Generate mock multi-wavelength galaxy catalog using EGG code}
We first build mock multi-wavelength galaxy catalogs using the Empirical Galaxy Generator \citep[EGG\footnote{https://cschreib.github.io/egg/},][]{2017A&A...602A..96S}, which is designed to generate mock galaxy catalog based on empirical prescriptions. The authors investigated stellar mass functions of quiescent galaxies (QGs) and SFGs (we refer them as passive and active galaxies hereafter in order to distinguish from radio AGN and SFG dichotomy in our simulation) selected according to UVJ criterion in the GOODS-south field. They draw stellar masses and redshifts from the (extrapolated) observed stellar mass functions of these two populations. They found close relationships between U-V colors and V-J colors in real observations which they called "UVJ sequence" and assigned these two colors for each simulated active and passive galaxy. They binned the UVJ plane and calculated the average SED of real galaxies in each bin, building an empirical library of 345 SEDs. Eventually, for a simulated active or passive galaxy with a certain stellar mass and redshift, its UVJ colors and stellar SED are progressively determined.

SFRs are assigned to each simulated active galaxy according to the SFR-$M_{\star}$ main sequence \citep[MS; see, e.g.,][]{2007ApJ...670..156D,2007ApJ...660L..43N,2011ApJ...739L..40R,2012ApJ...754L..29W,2015A&A...575A..74S,2017ApJ...847...76S} and 3\% of them are randomly chosen to be in "starburst" mode. For passive galaxies, the authors assigned them with residual SFR which they derived from stacking real galaxies. Once SFR is determined, it is decomposed into dust-unobscured component (emitting in the UV band) and dust obscured component (emitting in the IR band) with the ratio between IR luminosity and UV luminosity varying with stellar mass and redshift. To obtain FIR SEDs, the authors referred to a new SED library which relies on dust temperature ($T_{dust}$) and the $IR8=L_{IR}/L_{8}$ where $L_{8}$ is the luminosity at rest-frame 8 $\mu$m. They calibrated the redshift and stellar mass dependence of these two parameters \citep[see details in][]{2015A&A...575A..74S} and selected and rescaled the corresponding FIR SED after $T_{dust}$ and IR8 are determined for each simulated galaxy.

EGG has proven to be successful in reproducing the observed source counts from optical to far-IR bands. We use EGG code to generate a mock galaxy catalog covering a 2$\times$2 deg$^2$ area, selected as F444W < 28 mag. We choose F444W filter as selection band because this band is the reddest band that James Webb Space Telescope (JWST)-NIRcam will observe in the Cosmic Evolution Survey (COSMOS) field, which will be efficient in selecting high-z galaxies, a regime where SKA will largely contribute. There are 6,686,924 galaxies in the mock catalog with 6,234,820 and 452,104 of them being active and passive galaxies respectively. The redshift distribution spans a wide range until $z \sim 10$ and 98\% of them are below $z=6$. The maximum stellar mass reaches $10^{12}\ M_{\odot}$ with 63\% of them lying between $10^{8}-10^{11}\ M_{\odot}$.

\subsection{Radio AGN assignment}\label{assign}

Our next step is assigning radio AGNs in mock multi-wavelength source catalog. The most difficult issues involves decisions of AGN fractions and the relative abundance of AGNs with different radio luminosities. We take advantage of recent work by \citet{2024arXiv240104924W}, which gathered deep VLA 3 GHz observations on the Great Observatories Origins Deep Survey North (GOODS-N) and COSMOS fields. They used the ratio between IR luminosity and rest-frame 1.4 GHz radio luminosity, $q_{IR}$, defined as 

\begin{equation}
q_{IR}=\rm log  \frac{ \it L_{IR}}{\it L_{ 1.4\ GHz}}\times  3.75\times 10^{12}\ Hz
\end{equation}

to distinguish radio AGNs from normal SFGs, selecting $\sim$ 1000 radio AGNs in two fields. They carefully investigated AGN probability as a function of redshift, stellar mass and radio luminosity using maximum likelihood fitting and derived functional forms for radio AGNs in active and passive host galaxies respectively. The formulas are written as follows:

\begin{equation}
\begin{split}
p(L_{\rm 1.4\ GHz}\ |\ M_{\star},\ z)_{\rm active} &= \\
10^{-0.79} \left(\frac{M_{\star}}{M_0}\right)^{1.06} \left(\frac{L_{\rm 1.4\ GHz}}{L_0}\right)^{-0.77} \left(\frac{1+z}{1+z_0}\right)^{3.08}, \\
p(L_{\rm 1.4\ GHz}\ |\ M_{\star},\ z)_{\rm passive} &=\\
10^{-0.70} \left(\frac{M_{\star}}{M_0}\right)^{1.41} \left(\frac{L_{\rm 1.4\ GHz}}{L_0}\right)^{-0.60} \left(\frac{1+z}{1+z_0}\right)^{2.47}, \\
\label{eq1}
\end{split}
\end{equation}

For each stellar mass and redshift bin, we first set out to calculate the lower limit of 1.4 GHz radio luminosity above which we integrate the AGN probability in order to obtain the AGN fraction through

\begin{equation}
	f_{AGN}(M_{\star},z)|M_{\star},z=\int_{\rm lower\ limit}^{\infty }p(L_{\rm radio})\ |\ M_{\star},\ zdL_{\rm radio}
\label{eq2}
\end{equation}
We assume that the lower limit of 1.4 GHz radio luminosity is purely due to star formation activity, thus this question shifts to the measurements of the typical SFR for each stellar mass and redshift bin. We utilize star forming MS in \citet{2015A&A...575A..74S} to calculate $SFR_{MS}$ of each stellar mass and redshift bin. With $SFR_{MS}$ in hand, we then convert them into 1.4 GHz radio luminosities and regard them as the lower limit of the integration in Equation \ref{eq2}. We make use of the scaling relationship between SFR and 1.4 GHz radio luminosity in \citet{2003ApJ...586..794B}, 

\begin{equation}
SFR = 
\begin{cases}
    5.52\times10^{-22} L_{1.4\ GHz} ,\,\, L>L_{c}  \\
    \frac{5.52\times10^{-22}}{0.1+0.9(L/L_{c})^{0.3}}L_{1.4\ GHz} ,\,\,L\leq L_{c}
\end{cases}
\label{eq3}
\end{equation}
where $L_{c}=6.4\times 10^{21}$W Hz$^{-1}$ is the 1.4 GHz radio luminosity of a ~$L_{\star}$ galaxy. 

After $SFR_{MS}$, $L_{\rm limit\ 1.4\ GHz}$, $f_{AGN}|M_{\star},z$ in each stellar mass and redshift bin are progressively determined, we are able to calculate the total number of radio AGNs and randomly assign the corresponding number of mock galaxies that hosting radio AGNs. The next step is to determine the relative abundance of AGNs with different 1.4 GHz radio luminosities. We assume that they must follow the shape of RLF. \citet{2024arXiv240104924W} compared several forms of RLF of radio AGNs proposed in the literature and found that the pure density evolution (PDE) shape best fit their data. Here we use the same redshift-evolving shape to assign radio luminosities to these randomly chosen radio AGNs:
\begin{equation}
\Phi^{AGN}(L)=(1+z)^{\alpha_{D}}\times \frac{\Phi_{*,0}}{(L_{*,0}/L)^{\alpha}+(L_{*,0}/L)^{\beta}}
 \end{equation} 
 
where $\alpha_{D}$ follows $\alpha_{D}=-0.77\times z+2.69$, $\Phi_{*,0}=\frac{1}{0.5}10^{-5.5}\ \rm Mpc^{-3}\ dex^{-1}$ is the turnover normalization of the local RLF of radio AGNs, $L_{*,0}=10^{24.59}\ \rm W\ Hz^{-1}$ is the local turnover position, $\alpha=1.27$ and $\beta=-0.49$ are the slopes at the bright and faint end, respectively. For simplicity we divide 1.4 GHz luminosities into logarithm step of 0.1 ranging from $L_{limit\ 1.4\ GHz}$ to $10^{28}\ \rm W\ Hz^{-1}$ and assign the corresponding number of radio AGNs with their 1.4 GHz luminosity consistent with the weighted probability of each step $L_{1.4\ GHz}$ in every stellar mass and redshift bin. We note here that we only consider radio AGNs residing in host galaxies with stellar masses above $10^{10}\ M_{\odot}$ as radio AGNs tend to be hosted by massive galaxies \citep[e.g.,][]{2005MNRAS.362...25B}. Extrapolations into lower stellar mass regime in Formula \ref{eq1} will lead to excessive radio AGNs, especially at lower redshift (See Section \ref{def} for a detailed discussion).

\begin{figure*}[htbp]
    \centering
    \includegraphics[width=\linewidth]{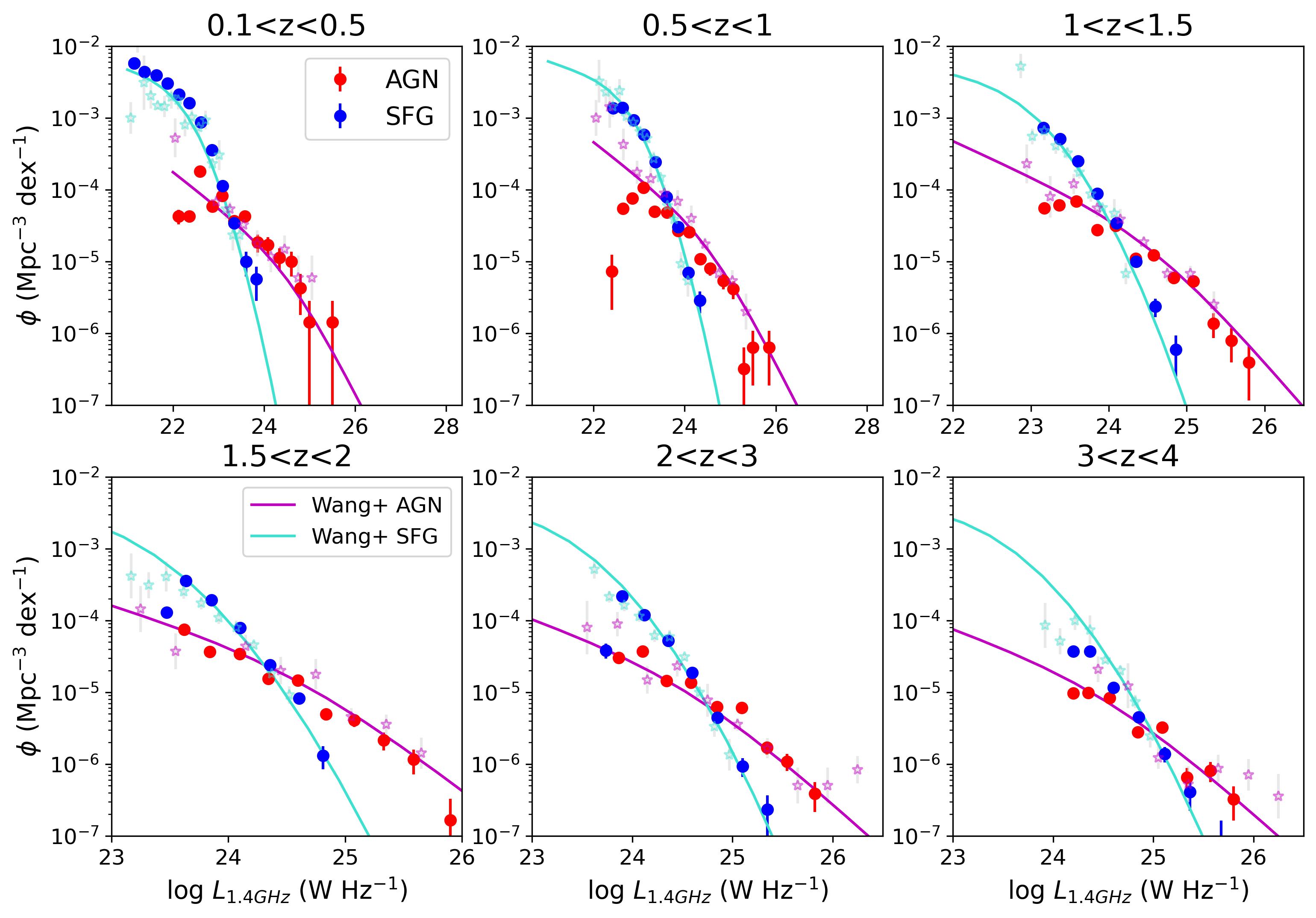}
\caption{Comparisons of RLFs of our mock AGNs (red) and SFGs (blue) to the results in \citet{2024arXiv240104924W} (magenta and cyan data points respectively), separated into different redshift bins. The magenta and cyan lines are best-fit RLF models in \citet{2024arXiv240104924W} for AGNs and SFGs respectively.}
\label{LF_wang}
\end{figure*}

\subsection{Radio luminosities of star forming galaxies}
After assigning radio AGNs based on AGN fraction $f_{AGN}\ |M_{*},\ z$ in each stellar mass and redshift bin, the remaining issue is to estimate 1.4 GHz radio luminosity for SFGs. We make use of $q_{IR}$ to convert IR luminosity into 1.4 GHz radio luminosity. We take advantage of $q_{IR}$ in \citet{2021A&A...647A.123D}, which calibrate $q_{IR}$ as a function of both stellar mass and redshift. The authors gathered > 400,000 star forming (active in our sense) galaxies in the COSMOS field and stacked available infrared/sub-mm and radio images in order to include sources that are undetected. For passive galaxies, there does not exist a well-defined scaling relationship between radio luminosities and IR luminosities, so we measure the distribution of their $q_{IR}$, irrespective of redshifts.

We make use of the "super deblended" far-IR (FIR) to sub-millimeter (sub-mm) photometric catalog in \citet{2018ApJ...864...56J}, which used the VLA-COSMOS 3 GHz \citep{2017A&A...602A...1S} and MIPS 24 $\mu$m detected sources as priors to deblend FIR and sub-mm images. They calculated IR luminosities via SED fitting. We cross-match the COSMOS2020 multi-wavelength catalog in \citet{2022ApJS..258...11W} with \citet{2018ApJ...864...56J} catalog within a matching radius of 0.5$\arcsec$ and find 890 3 GHz detected passive galaxies based on redshift-evolving UVJ criterion of passive galaxies in \citet{2011ApJ...735...86W}. We do not conduct stacking analysis for 3 GHz undetected sources due to poor S/N signal even after stacking 3 GHz images. We do not split them into redshift bins due to small number statistics. We calculate their $q_{IR}$ and find a peak value of 2.3.

For each EGG galaxy, we draw a $q_{IR}$ value from a Gaussian distribution with peak value corresponding to $q_{IR}$ in \citet{2021A&A...647A.123D} for active galaxies in each stellar mass and redshift bin and 2.3 for passive galaxies respectively. The standard deviation of each Gaussian distribution is set to be 0.26 dex \citep{2001ApJ...554..803Y}. Following \citet{2021A&A...647A.123D}, $q_{IR}$ values lower than 0.43 dex below the peak value of each Gaussian distribution is discarded as low $q_{IR}$ values are considered to be AGNs. With $q_{IR}$ in hand, we derive 1.4 GHz radio luminosities for SFGs.
\begin{figure*}[htbp]
    \centering
    \includegraphics[width=\linewidth]{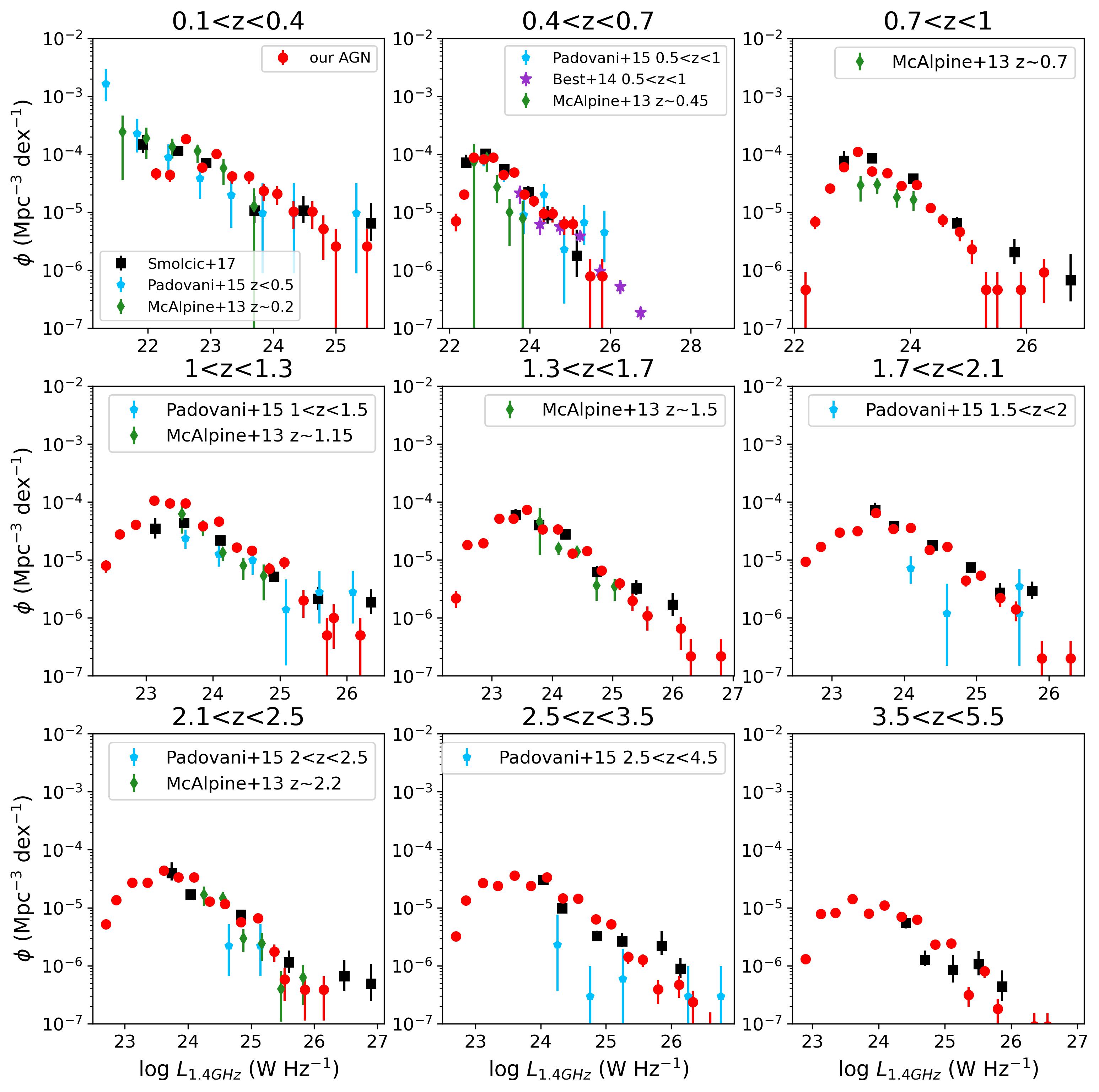}
\caption{Comparisons of RLFs of our mock AGNs (red) to literature data \citep{2013MNRAS.436.1084M,2015MNRAS.452.1263P,2017A&A...602A...6S}, separated into different redshift bins.}
\label{LF_agn}
\end{figure*}

\begin{figure*}[htbp]
    \centering
    \includegraphics[width=\linewidth]{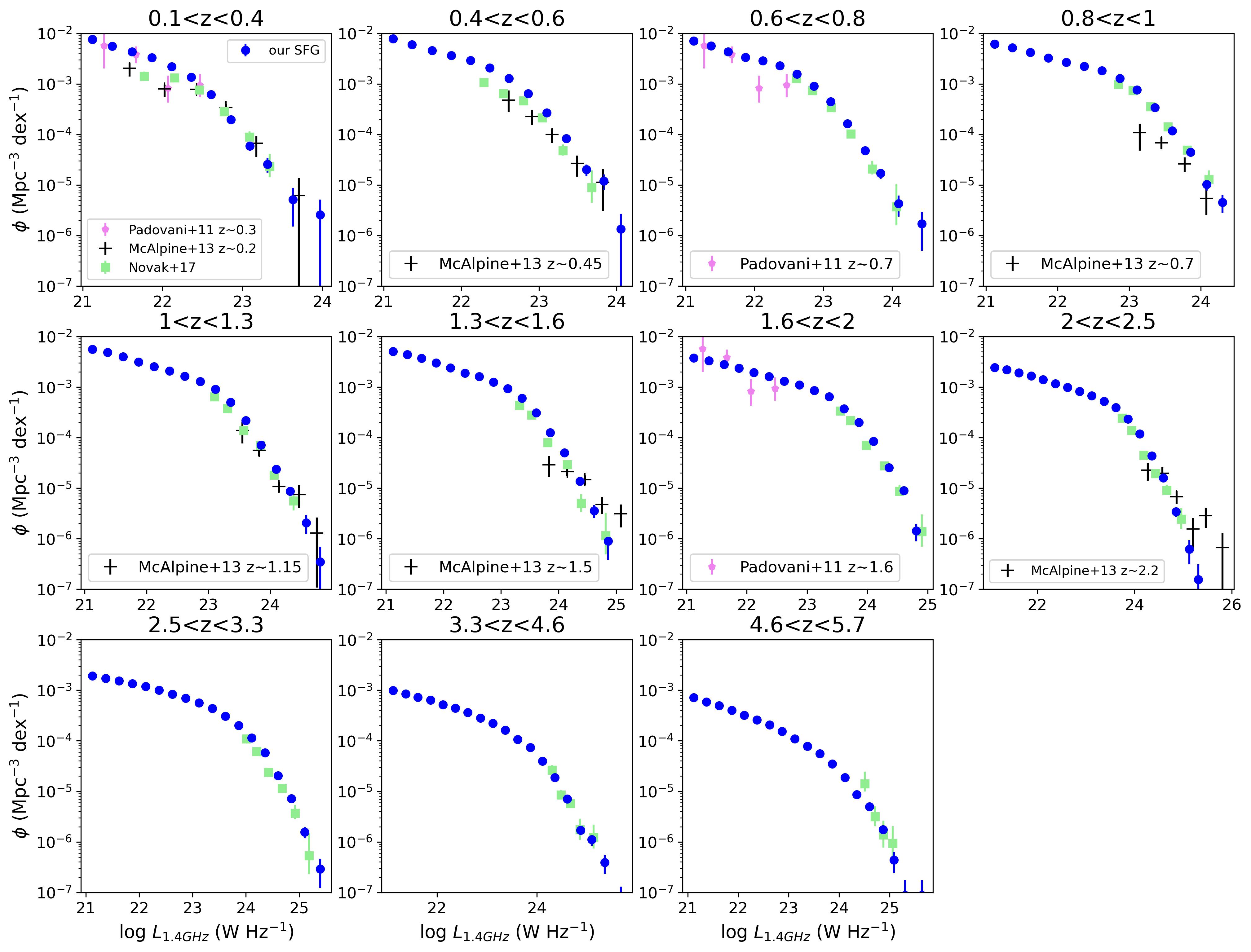}
\caption{Comparisons of RLFs of our mock SFGs (blue) to literature data \citep{2011ApJ...740...20P,2013MNRAS.436.1084M,2017A&A...602A...5N}, separated into different redshift bins.}
\label{LF_sfg}
\end{figure*}

\subsection{Clustering of radio AGNs}\label{clustering}
Radio sources are known to be more strongly clustered than optically selected galaxies, making them good indicators of large-scale structures \citep[see][and references therein]{2017MNRAS.464.3271M}. One popular way to measure the clustering signal of radio sources is through calculating the angular two-point correlation function (TPCF) $\omega(\theta)$, which quantifies the excess number of galaxies compared to random positions at different spatial scales. Literature works have revealed a power-law shape of angular TPCF of radio AGNs with a slope of $-0.8$. However, we only find weak clustering of randomly assigned mock radio AGNs. We argue that this is due to the weak clustering of EGG mock galaxies (see the left panel of Figure \ref{cluster}). Randomly assignment will smooth this clustering signal, indicating that we need to apply more strong clustering to their host galaxies.

Following \citet{2017A&A...602A..96S}, we redistribute EGG galaxies using the \citet{1978AJ.....83..845S} algorithm, which can reproduce an angular TPCF resembling real observations. We start from drawing a set of random positions in our 4 deg$^2$ simulated sky. A new set of $\eta$ positions is randomly drawn around each of the first set of random positions, within a radius of $R$. A third set of $\eta$ positions around the second level of $\eta$ positions is drawn, within a radius of $R/\lambda$. This procedure is repeated with reducing radius by a factor of $\lambda$ until we draw enough positions. We use the same parameters in \citet{2017A&A...602A..96S} (i.e., $R$ fixed to 3 \arcmin, $\eta=5$ and $\lambda=6$) to keep the same power law slope of $-1$ which massive galaxies follow as claimed by the authors. 

We place 500 random positions (first set of 3 \arcmin circles covering ~4 deg$^2$ in total) and repeat this procedure five times to obtain 321,500 positions, in order to cover all 256,373 EGG galaxies with stellar mass above $10^{10}\ M_{\odot}$. We testify this method by assigning different fractions of active and passive galaxies to these positions and generating random positions for the rest. Passive galaxies are known to display stronger clustering than active galaxies \citep[e.g.,][]{2008MNRAS.391.1301H,2010MNRAS.407.1212H}. We make sure that we reproduce this discrepancy between active and passive galaxies in our mock galaxies. After multiple trials, we find that assigning 15\% of active and 25\% of passive EGG galaxies respectively best reproduce clustering signal of these two populations, as shown in the middle panel of Figure \ref{cluster}. We demonstrate the result of angular TPCF of our mock radio AGNs in Section \ref{clusteringre}.

\begin{figure*}[htbp]
\centering
\begin{subfigure}{0.45\linewidth}
    \includegraphics[width=\linewidth]{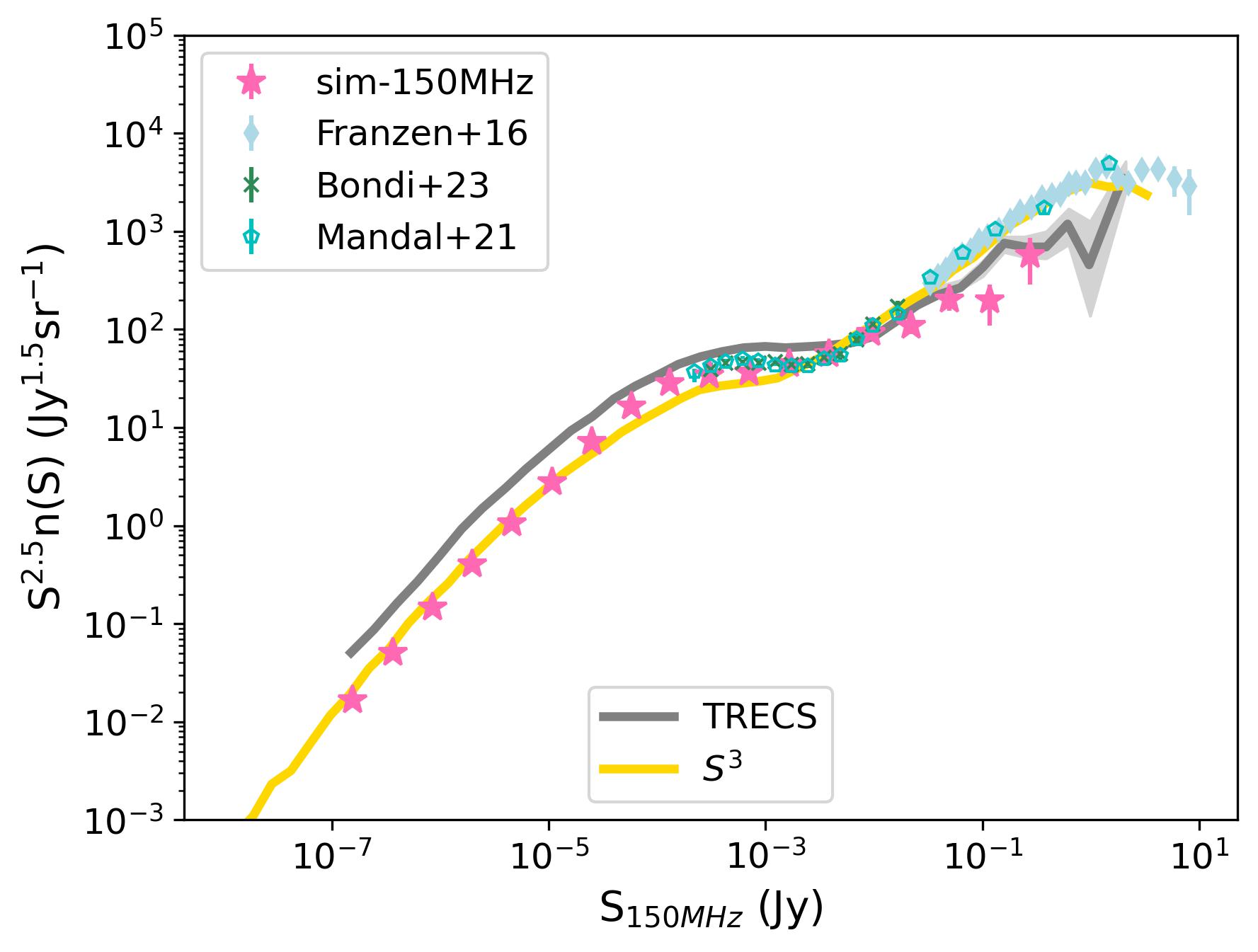}
    \caption{}
\end{subfigure}
\begin{subfigure}{0.45\linewidth}
    \includegraphics[width=\linewidth]{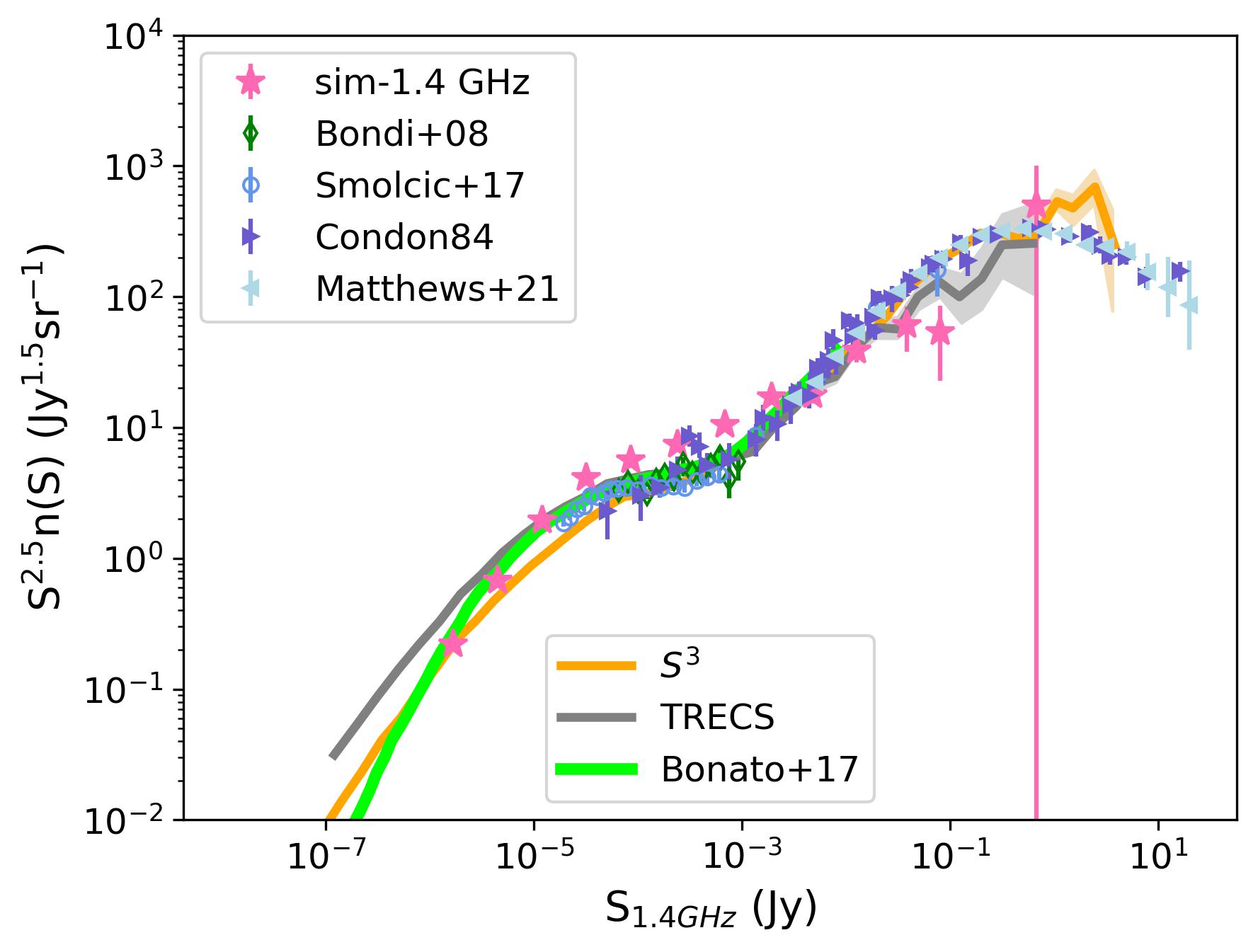}
    \caption{}
\end{subfigure}

\begin{subfigure}{0.45\linewidth}
    \includegraphics[width=\linewidth]{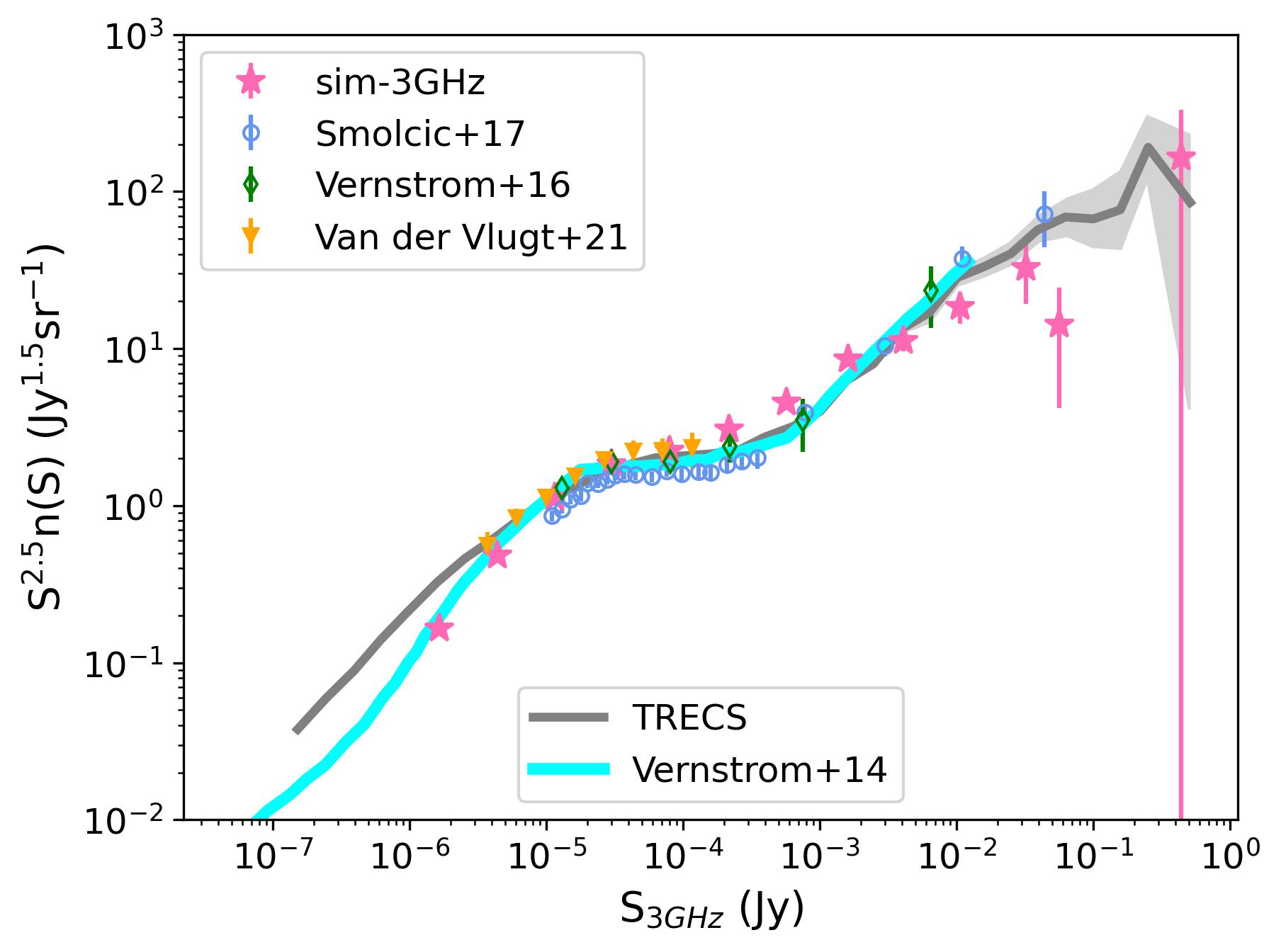}
    \caption{}
\end{subfigure}
\caption{\textit{a}: Comparisons of differential source counts at 150 MHz to observations in \citet{2016MNRAS.459.3314F,2021A&A...648A...5M,2023arXiv231206247B} along with $S^3$ and TRECS simulations \citep{2008MNRAS.388.1335W,2019MNRAS.482....2B}. \textit{b}: Comparisons of differential source counts at 1.4 GHz to observations in \citet{1984ApJ...284...44C,2008ApJ...681.1129B,2017A&A...602A...1S,2021ApJ...909..193M}, theoretical work in \citet{2017MNRAS.469.1912B}, along with $S^3$ and TRECS simulations \citep{2008MNRAS.388.1335W,2019MNRAS.482....2B}.\textit{c}: Comparisons of differential source counts at 3 GHz to observations in \citet{2016MNRAS.462.2934V,2017A&A...602A...1S,2021ApJ...907....5V}, $P(D)$ analysis in \citet{2014MNRAS.440.2791V} and TRECS simulation \citep{2019MNRAS.482....2B}.}
\label{count}
\end{figure*}

\subsection{Physical sizes of SFGs}
Galaxy sizes are necessary in producing realistic images. We assume that radio-loud AGNs are point-like sources and only account for simulating physical sizes for SFGs. We take advantage of the stellar-mass-dependent form of galaxy effective radius provided in \citet{2014ApJ...788...28V}:
\begin{equation}
R_{\rm eff} (\rm kpc)=A( M_{\star}/5\times10^{10}M_{\odot})^{\alpha}
\label{size}
\end{equation}
with A and $\alpha$ being free parameters. To obtain best-fit values for these free parameters we apply Formula \ref{size} to radio SFGs in the GOODS-S fields as VLA 3 GHz observations in this field reach unprecedented depth of 0.75 $\mu$Jy beam$^{-1}$ and resolution of 0{\arcsec}.6 $\times$ 1{\arcsec}.2 \citep{2020ApJ...901..168A}. We cross match the blind-extracted radio catalog with the multi-wavelength source catalog of \citet{2013ApJS..207...24G} within 2{\arcsec} radius and find 385 sources with peak 3 GHz flux densities above 5$\sigma$. We run source extractions via \texttt{pybdsf} in the 3GHz image to obtain the FWHM of the deconvolved major axis then divided by 2.43 to derive the effective radius according to \citet{2017ApJ...839...35M}. We retrieve 358 radio sources among which 115 sources are considered as SFGs according to the demarcation lines between radio AGNs and SFGs in Section \ref{dem} and have signal-to-noise ratio above 10. We derive the best-fit values of A and $\alpha$ based on Markov chain Monte Carlo method via \texttt{emcee} in three redshift bins, in order to take account into any redshift-dependent evolution. The values are listed in Table \ref{R_para}. We then fit the discrepancy between observed effective radius and model values with a Gaussian distribution centered on zero and obtain a standard deviation of 0.95 kpc. For each of our mock SFG, we calculate their effective radius based on their stellar masses according to Formula \ref{size} and add a random value draw from a Gaussian distribution with mean and dispersion equaling to 0 and 0.95 kpc respectively. We force the physical size should not be smaller than 0 kpc or larger than 10 kpc.

\begin{table}[]
    \centering
    \caption{Best-fit parameters of effective radius detected on GOODS-S 3 GHz images.}
    \label{R_para}
    \begin{tabular}{c c c c}
    \hline\hline
      redshift bins&number of SFGs   &A & $\alpha$\\
      \hline
        0<z<1 &56 &2.093$^{+0.180}_{-0.170}$&0.092$^{+0.034}_{-0.039}$\\
        1<z<2 & 40&2.415$^{+0.089}_{-0.086}$&0.029$^{+0.027}_{-0.020}$\\
         z>2&19 &1.737$^{+0.143}_{-0.132}$&0.211$^{+0.075}_{-0.080}$\\
         \hline
    \end{tabular}
\end{table}

\section{Validations}\label{validation}
In this section, we compare our mock radio sources with literature work. We first compare the rest-frame 1.4 GHz RLF of AGNs and SFGs respectively, then we compare the normalized source counts at three frequencies, namely 3 GHz, 1.4 GHz, and 150 MHz with observational and theoretical work. Finally, we calculate the angular TPCF of radio AGNs, after redistributing EGG galaxies and assigning radio AGNs. We use a median spectral slope of $\alpha=-0.8$ to derive flux densities of SFGs at 3 GHz and 1.4 GHz. At 150 MHz, we use a median spectral slope of $\alpha=-0.5$ according to a recent work by \citet{2023arXiv230306941A}. The authors selected ~3500 SFGs that are detected by both the Low Frequency Array (LOFAR) at 150 MHz and the Giant Metrewave Radio Telescope (GMRT) at 610 MHz in the ELAIS-N1 field and found a peak value of $\alpha=-0.5$. For radio AGNs, we adopt a median slope of $\alpha=-0.5$ which is a value between a flat ($\alpha \sim 0$) and a steep ($\alpha=-0.8$) spectrum \citep[e.g.,][]{1983ApJ...269..352S,1989AJ.....98.1195K,2007ApJ...656..680J}. To mimic variations in spectral slopes, for each source we first generate a random alpha slope from a gaussian distribution with peak value equaling to either $-0.8$ or $-0.5$ and standard deviation of 0.1. Then we derive the flux densities at each band accordingly.

\subsection{Radio luminosity functions}\label{LF}
Fig. \ref{LF_wang} compares the RLF of our mock AGNs (red dots) and SFGs (blue dots) to the data points (magenta and cyan stars respectively) in \citet{2024arXiv240104924W}. Their best-fit models are shown in magenta and cyan solid lines, respectively. We first select radio sources with 3 GHz flux density above 2.3 $\mu$Jy to mimic real COSMOS-VLA observations \citep[see][since we focus on one-beam dominated sources we regard $\mu$Jy as the same as $\mu$Jy beam$^{-1}$]{2017A&A...602A...1S}, then we apply $V_{max}$ corrections to calculate RLFs in each 1.4 GHz radio luminosity bin via 
\begin{equation}
\Phi(L_{1.4 GHz},z)=\frac{1}{\Delta log L_{1.4GHz}}\sum_{i}\frac{1}{\frac{\Omega}{4\pi} \times V_{max,i}}
\end{equation}
where $V_{max,i}=V_{zmax,i}-V_{zmin,i}$, with $V_{zmax,i}$ being the co-moving volume at the maximum redshift where the $i$\ th source can be observed given the detection limit, and $V_{zmin,i}$ being the co-moving volume at the lower boundary of the corresponding redshift bin. We find a good agreement between our mock radio sources and their results until z $\sim$ 4 for both populations.

In Fig. \ref{LF_agn}, we compare the RLF of our mock AGNs (red dots) with several literature works. We do not apply any selection cut this time in order to explore the lower radio luminosity regime. \citet{2013MNRAS.436.1084M} selected 1054 radio sources detected at 1.4 GHz from the VIRMOS VLT Deep Survey \citep[VVDS,][]{2003A&A...403..857B} covering 1 deg$^2$, reaching a 5$\sigma$ limit of 80 $\mu$Jy. \citet{2015MNRAS.452.1263P} selected 765 radio sources covering 0.285 deg$^2$ in the Extended Chandra Deep Field-South (E-CDFS) with 1.4 GHz flux density $\geq 32.5\ \mu$Jy. \citet{2017A&A...602A...6S} made use of the deep VLA-COSMOS data, containing more than 1800 radio AGNs spanning a wide redshift range out to $z \sim 5$. We find consistent results with literature works at several redshift bins in overlapped luminosity range. On the other hand, unlike previous studies which draw sources based on extrapolated RLFs \citep[e.g.,][]{2019MNRAS.482....2B}, we find reduced 1.4 GHz RLF at low luminosity end above $z=0.4$, indicating fewer less luminous radio AGNs at early epochs than simple extrapolations. This is reasonable given the fact that fewer galaxies can grow into massive galaxies with $M_{\star}>10^{10}$ in the early universe compared to in the local universe. In addition, $SFR_{MS}$ is higher towards higher redshifts which results in larger $L_{limit\ 1.4\ GHz}$. Therefore, the RLF cannot extrapolate to very faint regime at high redshifts. Currently there is no available observational data probing this fainter regime, more powerful next-generation facilities (such as SKA) may help unveil this mystery.

In Fig. \ref{LF_sfg} we compare the RLF of our mock SFGs (blue dots) to other literature data. Again we do not apply any flux density selection criterion. \citet{2011ApJ...740...20P} studied a sample of 256 1.4 GHz selected sources in the CDFS field, reaching a flux density limit of 43 $\mu$Jy at the field center. A more recent work by \citet{2017A&A...602A...5N} took advantage of the deep VLA-COSMOS 3 GHz observations and selected $\sim 6000$ SFGs spanning a wide redshift range out $z\sim 5$ across 2 deg$^2$. We also find excellent agreement with real observations, even at low luminosity end until $z \sim 2$. Still, more sensitive data is needed to testify the low luminosity end of the RLF towards higher redshifts.

\begin{figure*}[htbp]
\begin{subfigure}{0.3\textwidth}
    \centering
    \includegraphics[width=\linewidth]{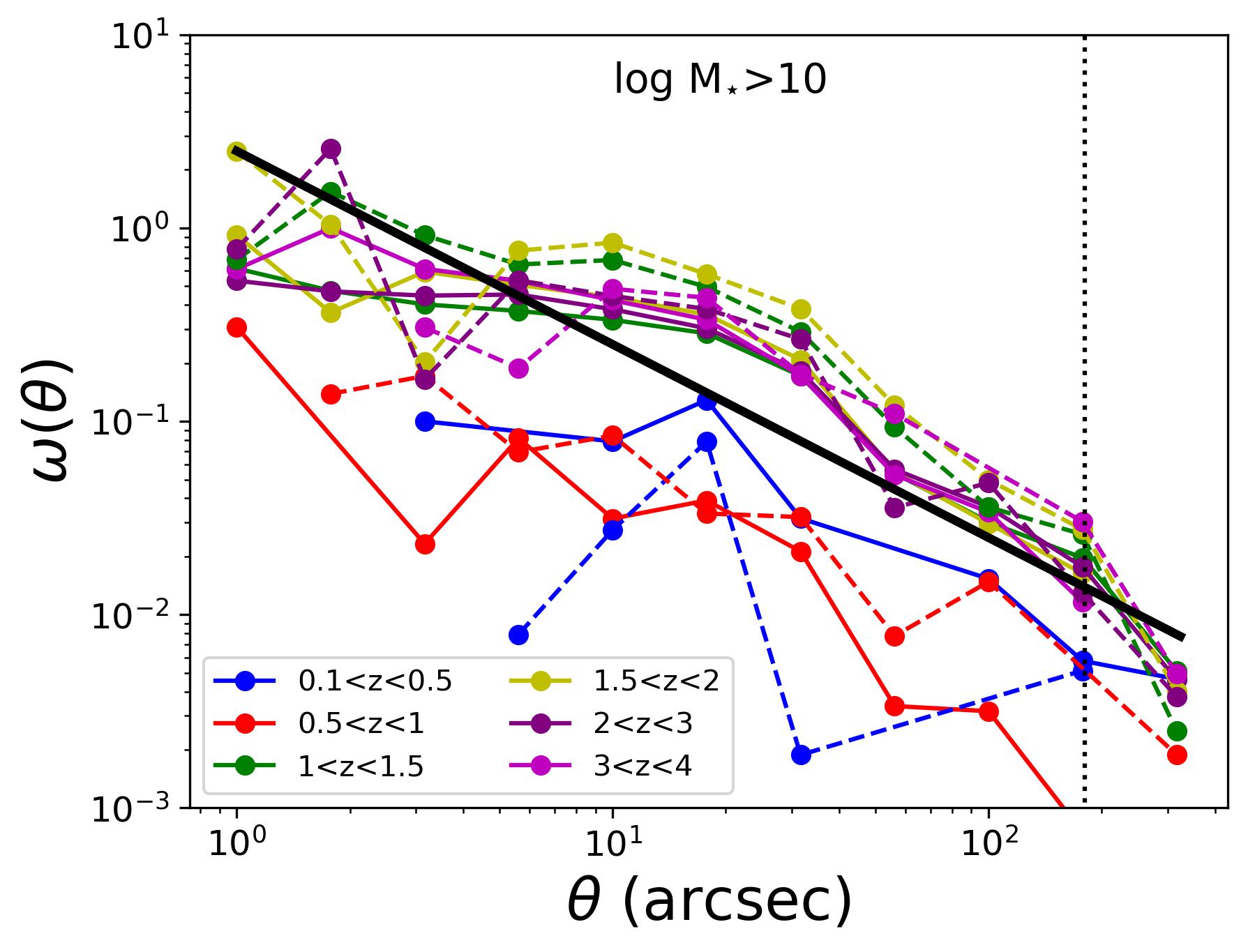}
\end{subfigure}
\begin{subfigure}{0.3\textwidth}
    \centering
    \includegraphics[width=\linewidth]{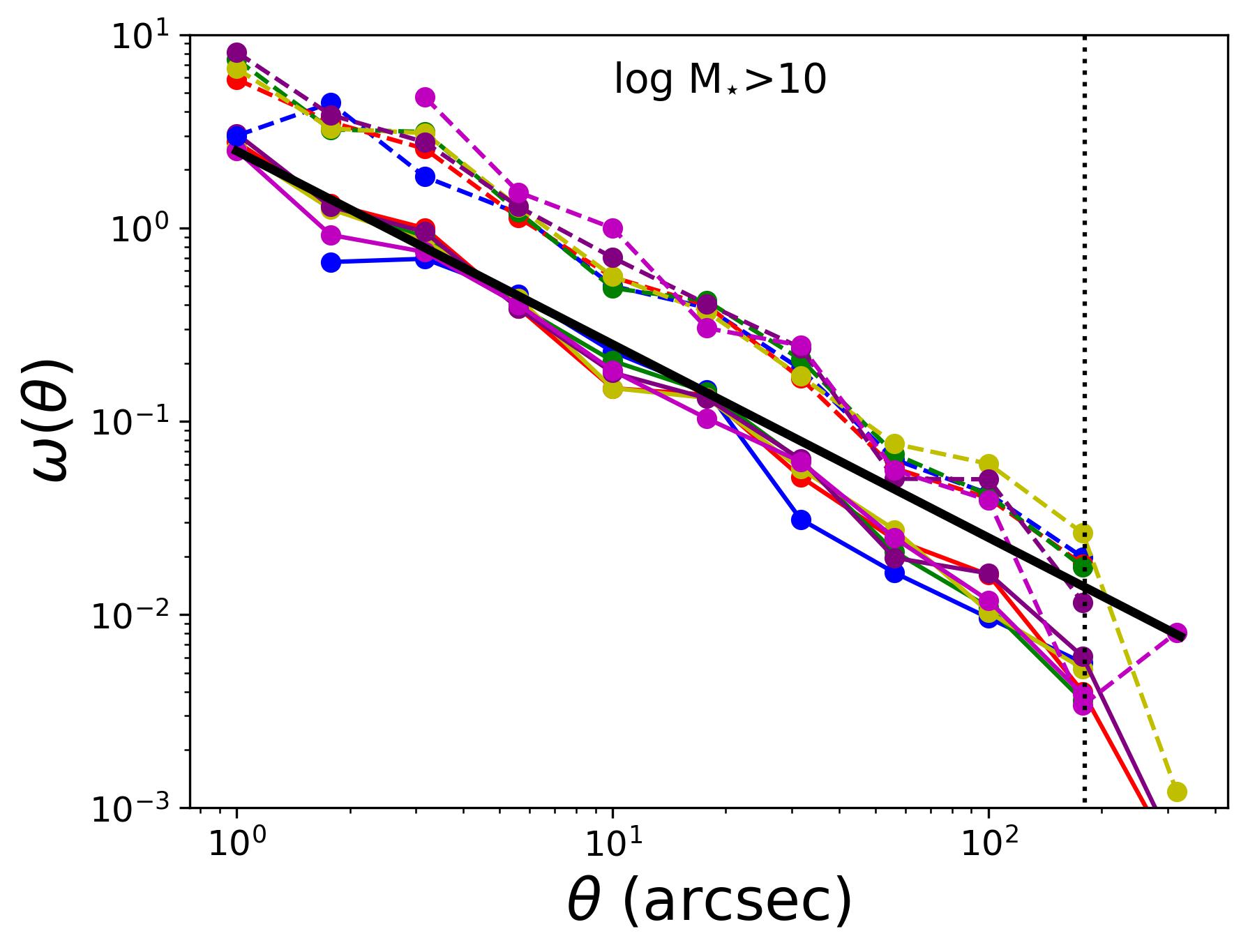}
\end{subfigure}
\begin{subfigure}{0.3\textwidth}
    \centering
    \includegraphics[width=\linewidth]{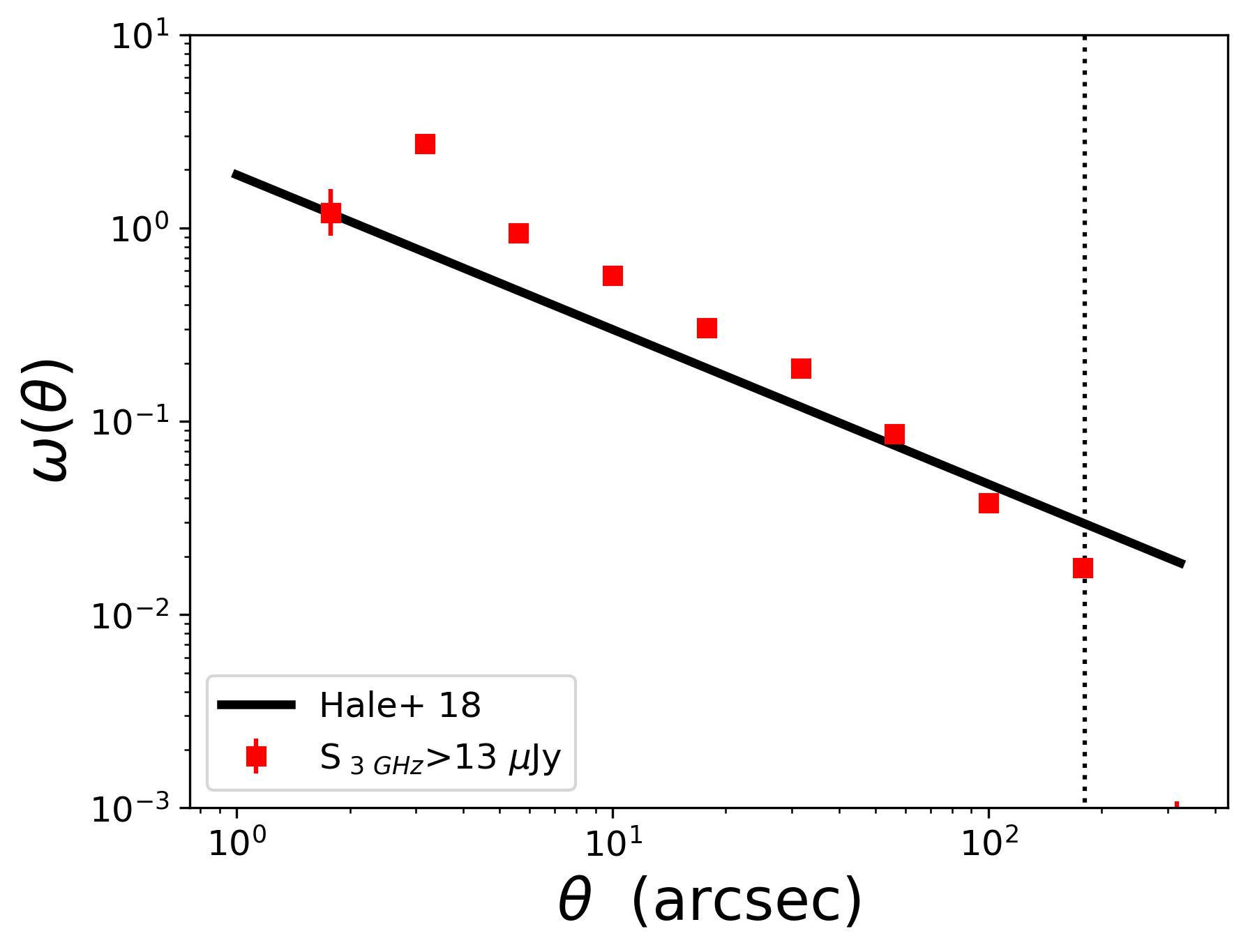}
\end{subfigure}

\caption{\textit{left}: Angular TPCF of mock EGG galaxies with stellar mass $>10^{10}\ M_{\odot}$ at different redshift bins. A straight line with power law slope of $-1$ is also plotted. \textit{middle:} Angular TPCF of redistributed mock EGG galaxies with stellar mass $>10^{10}\ M_{\odot}$ at different redshift bins. A straight line with power law slope of $-1$ is also plotted. Solid and dashed lines represent active and passive galaxies respectively. \textit{right:} Angular TPCF of our mock radio AGNs with 3 GHz flux density above 13 $\mu$Jy. We also plot the best-fit model from \citet{2018MNRAS.474.4133H} with a slope of $-0.8$. The vertical dotted line denotes the clump radius of 3{\arcmin} (see Section \ref{clustering}).}
\label{cluster}
\end{figure*}

\subsection{Differential source counts}\label{counts}
In Fig. \ref{count} we compare the normalized source counts of our mock radio sources at 150 MHz, 1.4 GHz, and 3 GHz to various observational data and model predictions. Early study in \citet{1984ApJ...284...44C} combined several 1.4 GHz surveys of three different instruments, namely the National Radio Astronomy Observatory (NRAO) 91m telescope for sources stronger than 2 Jy, VLA and Westerbork. \citet{2008ApJ...681.1129B} investigated the early release of COSMOS-VLA 1.4 GHz project with a resolution of 1.5{\arcsec} and a sensitivity of $\sim 11\ \mu$Jy, yielding $\sim$ 3600 radio sources. \citet{2016MNRAS.462.2934V} studied 558 sources detected above 5$\sigma$ at 3 GHz in the Lockman Hole North field, with a instrument noise of 1.01 $\mu \rm Jy\ beam^{-1}$. \citet{2017A&A...602A...1S} took advantage of the deeper VLA-COSMOS data, providing 10830 radio sources detected above 5$\sigma$ at 3 GHz, reaching a median rms of 2.3 $\mu \rm Jy\ beam^{-1}$. More recently, \citet{2021ApJ...909..193M} measured 1.4 GHz source counts based on MeerKAT DEEP2 images for flux densities below 2.5 mJy and NRAO VLA Sky Survey (NVSS) for flux densities above 2.5 mJy. \citet{2021ApJ...907....5V} present ultra deep VLA 3 GHz observations in the COSMOS field, reaching median rms of 0.53 $\mu \rm Jy\ beam^{-1}$ in a total area of 180 arcmin$^2$.

At low frequency, \citet{2016MNRAS.459.3314F} studied 154 MHz images using the Murchison Widefield Array (MWA), covering a large sky area of 570 deg$^2$ with rms noise of 4--5 mJy beam$
^{-1}$. With the advent of more powerful low frequency facility LOFAR, \citet{2021A&A...648A...5M} and \citet{2023arXiv231206247B} were able to push the 154 MHz flux density limit down to a few tens of $\mu$Jy beam$^{-1}$, in combined ELAIS-N1, Lockman Hole and Bo$\rm \ddot{o}$tes fields and the Euclid Deep Field North (EDFN) field respectively.

We find consistent results with observational data for all three frequencies. Towards lower flux densities, our source counts are more consistent with confusion amplitude distribution $P(D)$ analysis in \citet{2014MNRAS.440.2791V} at 3 GHz and theoretical models in \citet{2017MNRAS.469.1912B} at 1.4 GHz respectively. $P(D)$ is the probability distribution of peak flux densities in an image. $P(D)$ analysis can allow for a statistical estimate of source counts below the confusion limit of surveys. \citet{2014MNRAS.440.2791V} fitted their $P(D)$ models to 3 GHz observations in the Lockman Hole field and managed to constrain the source counts down to 50 nJy, a factor of 20 below the rms confusion. \citet{2017MNRAS.469.1912B} derived source counts using empirical relations between SFR and free-free emission and synchrotron luminosity for SFGs, along with RLF models for AGNs from \citet{2010MNRAS.404..532M}. This good agreement between our work and their results suggests the feasibility of using $q_{IR}$ of SFGs to derive radio source counts at fainter end, once the IR luminosities of SFGs are known. Still, more powerful next-generation radio facilities are needed to further investigate this faint regime.





\subsection{Clustering of radio AGNs}\label{clusteringre}
We verify the clustering of our mock radio AGNs by calculating the angular TPCF $\omega(\theta)$. As in \citep{2019MNRAS.482....2B}, we adopt the \citet{1993ApJ...417...19H} estimator: \[
	\omega(\theta)=\frac{DD\times RR}{DR \times DR}-1
\]
where DD, DR and RR are the number of galaxy pairs between data and data samples, data and random samples, and random and random samples respectively.

\citet{2018MNRAS.474.4133H} measured the angular TPCF of the deep 3 GHz detected sources in the COSMOS field. They classified AGNs based on a wide range of criteria such as X-ray luminosity, mid-IR color and so on \citep[see][]{2017A&A...602A...2S}. They used a 5$\sigma$ flux density limit of $\sim 13\ \mu$Jy at 3 GHz and fitted the observed TPCF with a power law slope of $-0.8$. We select a total of 6274 radio AGNs with 3 GHz flux density above $13\ \mu$Jy. We randomly place ten time more sources across the same 4 deg$^2$ area and count the DD, DR and RR galaxy pairs. We repeat this procedure 200 times and use the median and 25/75 percentiles of 200 outputs as final $\omega(\theta)$ values. In the right panel of Fig. \ref{cluster} we plot the results and compare to the best-fit model in \citet{2018MNRAS.474.4133H}. After redistributing EGG massive galaxies and randomly assigning radio AGNs, we observe strong TPCF amplitudes at small spacial scales for radio AGNs. However, we cannot reproduce TPCF amplitudes towards larger spacial scales as the maximum scale radius used in \citet{1978AJ.....83..845S} algorithm in Section \ref{clustering} only reaches 3\arcmin. In addition, the final slope of TPCF is more close to $-1$ as we adopt the same parameters of the \citet{1978AJ.....83..845S} algorithm as used in EGG code, which they claimed that this combination of parameters can reproduce the observed $\alpha=-1$ slope of TPCF of massive galaxies. Despite of these deficiencies, our work is still able to reproduce the clustering signal of radio AGNs without introducing cosmological simulation of dark matter halos.


\section{Applications of our simulation in the era of SKA and LSST}\label{app}
\subsection{What type of galaxies can (only) be observed by SKA?}
\begin{figure*}[htbp]
    \centering
    \includegraphics[width=\linewidth]{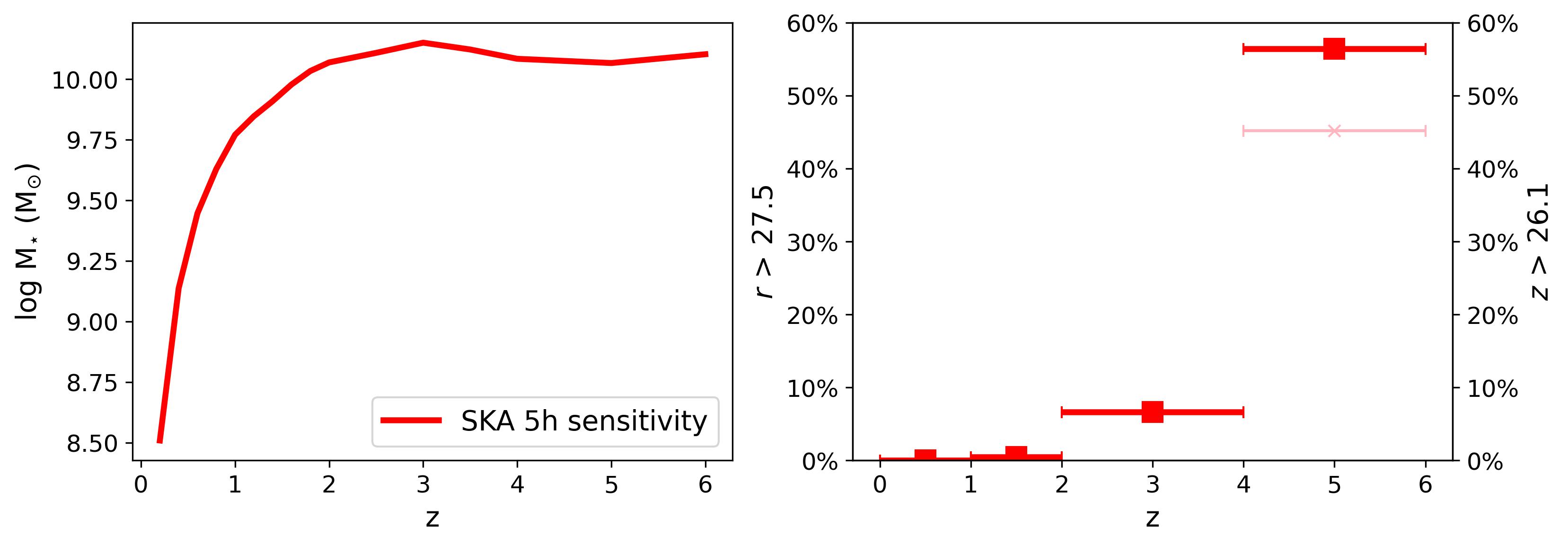}
    \caption{\textit{left:} The 90\% completeness of stellar mass of galaxies that can be observed above 5$\sigma$ by SKA as a function of redshift, assuming 5 hour observation time for a single pointing.  \textit{right:} Red squares denote the fraction of SKA detected galaxies that are above 27.5 mag in the SDSS-$r$ band (approximate to LSST-$r$ band depth) at different redshifts. At $4<z<6$, pink cross represents the fraction of SKA detected galaxies that are above 26.1 mag in the SDSS-$z$ band (aproximate to LSST-$z$ band depth). }
    \label{mass_z}
\end{figure*}

In the left panel of Fig. \ref{mass_z} we display the 90\% completeness stellar mass of galaxies that can be observed by SKA as a function of redshift. We select SFGs with 1.4 GHz flux density above five times of 0.447 $\rm \mu Jy\, beam^{-1}$, which is drawn from the SKA Sensitivity Calculator\footnote{https://sensitivity-calculator.skao.int/mid. We adopt default AA4 configuration at Band 2, centered at 1.4 GHz with a bandwidth of 500 MHz under natural weighting.} assuming 5 hours observational time for a single pointing (the time spent for each VLA pointing). We are expected to be able to identify many low mass galaxies at higher redshifts, $\sim 10^{10}$ M$_{\odot}$ up to $z \sim 6$, in the SKA era, significantly benefiting the study of galaxy formation in the early universe.

In addition to the stellar mass range that can be observed by SKA, we would like to probe how many these SKA-detected galaxies may also be identified by large sky survey in the optical band. The Large Synoptic Survey Telescope (LSST), currently known as the Vera C. Rubin Observatory, will carry out an ambitious survey in the optical band using its 8.4-meter primary mirror. The planned 800 exposures over 10-years observations are expected to reach a depth of $r \sim 27.5$ \citep{2019ApJ...873..111I}. In the right panel of Fig. \ref{mass_z} we approximate the LSST-$r$ band with the SDSS-$r$ band and illustrate the fractions of galaxies that are fainter than 27.5 mag among those can be detected by SKA. We find an increasing fraction of SKA-detected galaxies that will be missed in deep optical surveys towards early epochs, reaching 56\% at $z>4$. For sanity check we also show the fraction of SKA detected sources that will be missed by longer $z$ band observations of LSST \citep[having a 5$\sigma$ depth of 26.1 mag;][]{2019ApJ...873..111I}. There still exist a fraction of 45\% of SKA detected sources that will appear invisible in the deep $z$ band image.
This plot demonstrates that SKA will play a critical role in uncovering massive dust-obscured galaxies, in complement with deep optical survey at higher redshifts. We refer to Section \ref{z4caution} for a detailed discussion on the robustness of our simulation at $z>4$.

\begin{figure*}[htbp]
\begin{subfigure}{0.46\textwidth}
    \centering
    \includegraphics[width=\linewidth]{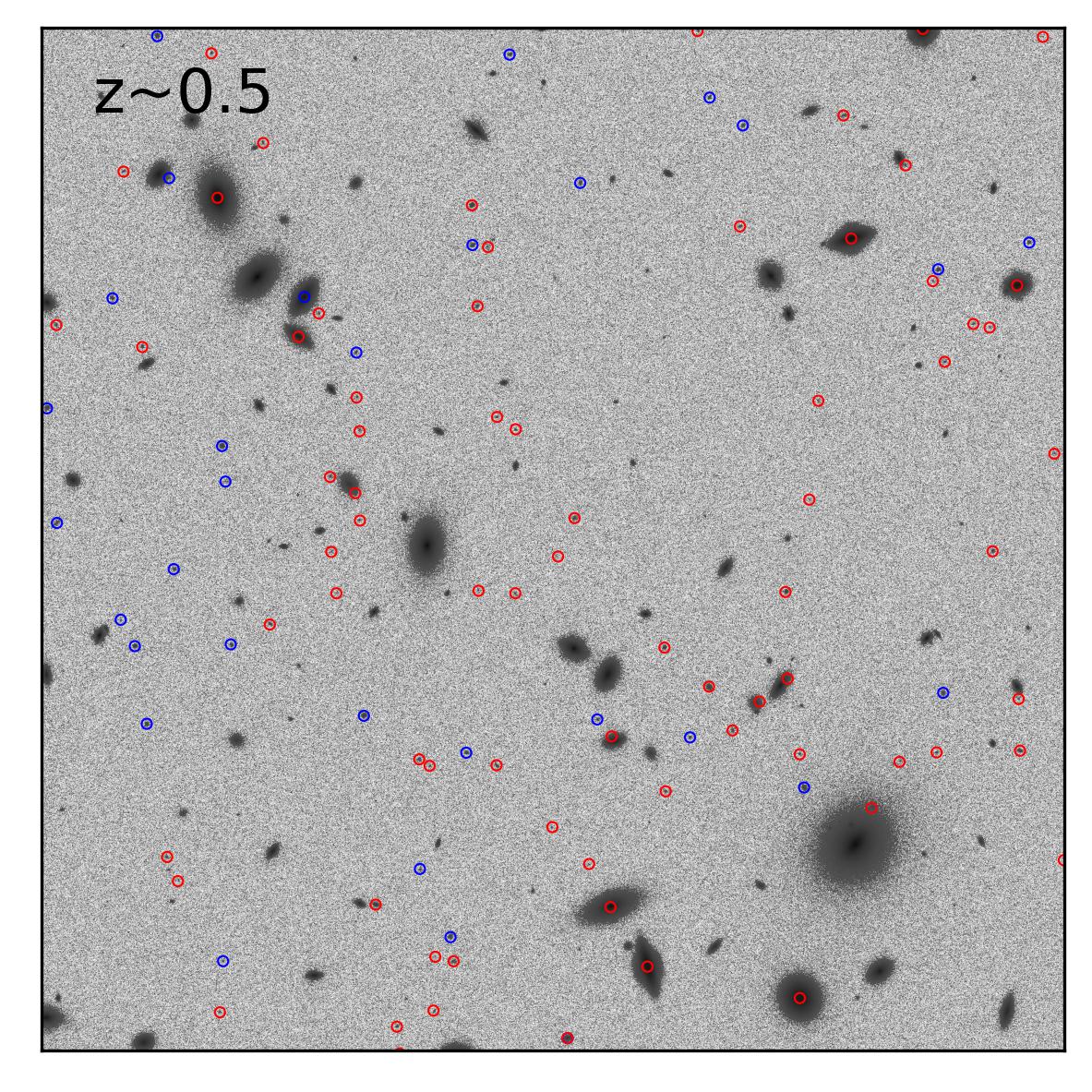}
\end{subfigure}
\begin{subfigure}{0.46\textwidth}
    \centering
    \includegraphics[width=\linewidth]{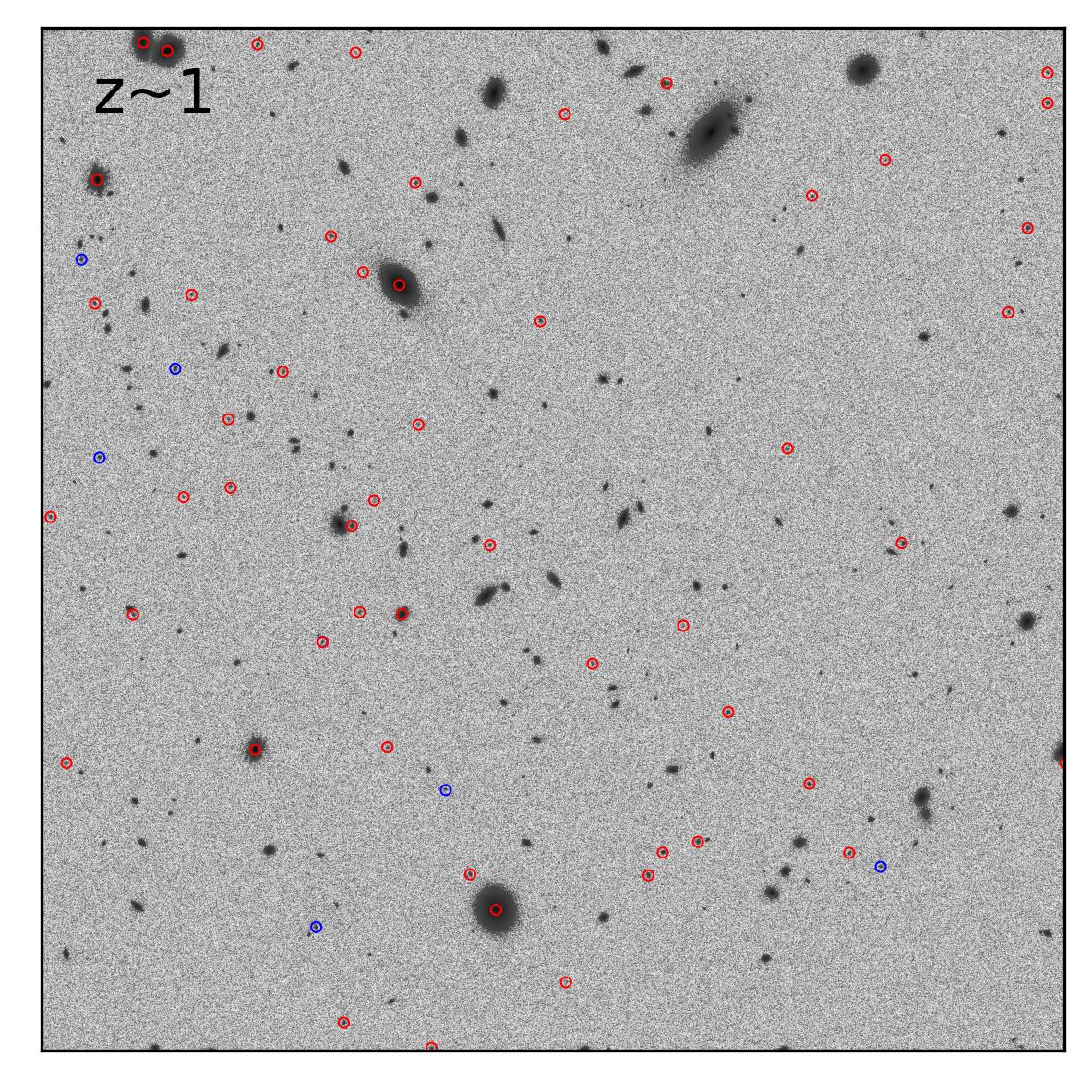}
\end{subfigure}

\begin{subfigure}{0.46\textwidth}
    \centering
    \includegraphics[width=\linewidth]{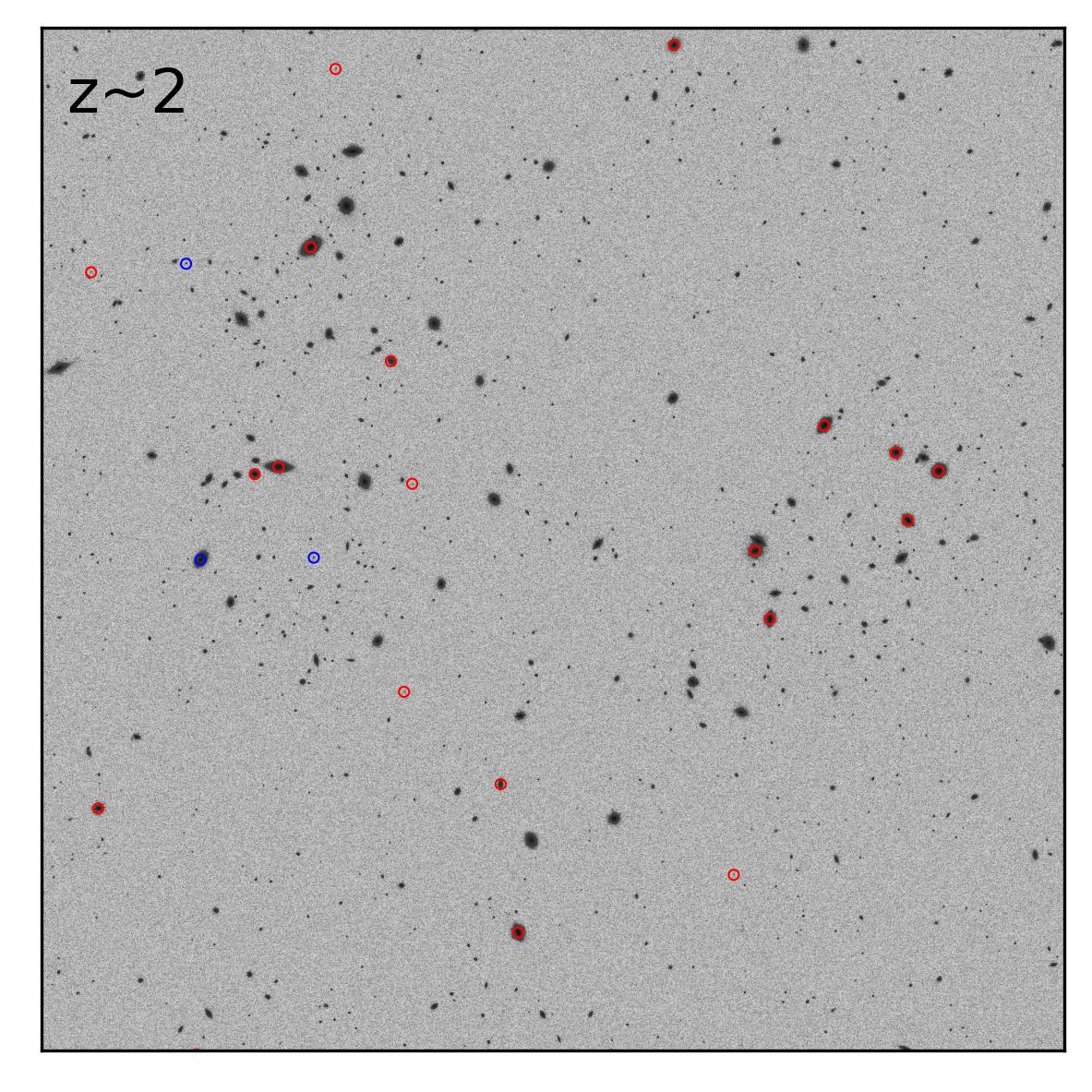}
\end{subfigure}
\begin{subfigure}{0.46\textwidth}
    \centering
    \includegraphics[width=\linewidth]{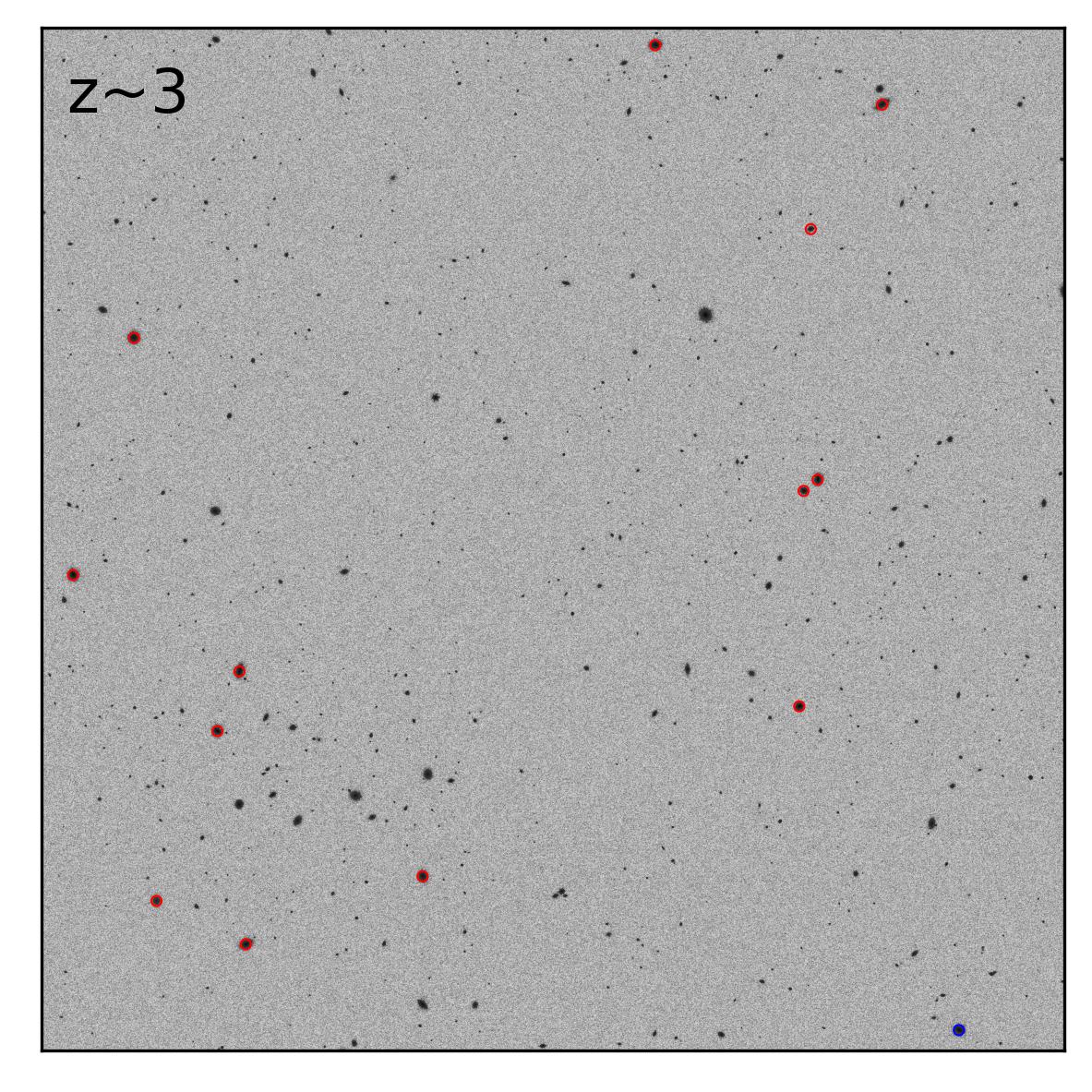}
\end{subfigure}

\caption{Top: Simulated $2\arcmin \times 2 \arcmin$ JWST-F444W images at $0.25<z<0.75$ and $0.75<z<1.25$. Bottom: Simulated $4\arcmin \times 4 \arcmin$ JWST-F444W images at $1.75<z<2.25$ and $2.75<z<3.25$. Blue circles represents sources detected under VLA observation (with rms noise level of 4.23 $\rm \mu Jy\, beam^{-1}$ at 1.4 GHz) and red circles represents sources only revealed by SKA observation (with rms noise level of 0.447 $\rm \mu Jy\, beam^{-1}$) at 1.4 GHz.}
\label{images}
\end{figure*}

In order to have a closer look of the optical morphology of SKA detected galaxies, we simulate radio images and run source extractions. To simulate radio images we assume that radio AGNs occupy a single pixel \citep[using pixel size of 0.24\arcsec, see][]{2021MNRAS.500.3821B} with their flux density equaling to peak flux density. On the other hand, we assume SFGs showing exponential (S$\acute{\rm e}$rsic profile with $n=1$) profile with their simulated flux density equaling to total flux densities. Following \citet{2016MNRAS.463.3339T}, we first generate the absolute ellipticity (|E|) for SFGs through its distribution :
\begin{equation}
    P(|E|)=|E|cos^2\frac{\pi|E|}{2}exp(-\frac{2|E|}{B})^C
\end{equation}
with B=0.19 and C=0.58. \citet{2016MNRAS.463.3339T} considered a two-component ellipticity, $\textbf{E}=(E_1,E_2)$, where the first component describing elongations parallel and perpendicular to a chosen reference axis while the second component describing elongations along the directions rotated $\pm 45\degree$ from the reference axis. The ellipticity modulus $E=|E|$ is connected to widely-used ellipticity $e=b/a$ through 
 \begin{equation}
     |E|=\frac{1/e^2-1}{1/e^2+1}.
 \end{equation}
The position angles are drawn from an uniform distribution between 0 and 2$\pi$. After implementing radio AGNs and SFGs , we conduct point-spread-function (PSF) convolution \citep[using PSF size of 0.6\arcsec, see][]{2021MNRAS.500.3821B} and add Gaussian noise. As above we run two trials with Gaussian noise of 4.23 $\rm \mu Jy\, beam^{-1}$ and 0.447 $\rm \mu Jy\, beam^{-1}$ (0.6 $\rm \mu Jy\, arcsec^{-2}$ and 0.06 $\rm \mu Jy\, arcsec^{-2}$ respectively assuming beam size equaling to PSF size) which correspond to VLA and SKA observations under the same 5 hour observation per pointing respectively. We run \texttt{pybdsf} for these two mock 1.4 GHz images and present simulated JWST-F444W images with detected radio sources at four redshift bins in Fig. \ref{images}. Blue circles represent radio sources that are detected by VLA and red circles denote radio sources that are only revealed by SKA. The unparalleled sensitivity of SKA will help us detect many more extragalactic radio sources.

\begin{figure*}[htbp]
\centering
\begin{subfigure}{0.45\linewidth}
    \includegraphics[width=\linewidth]{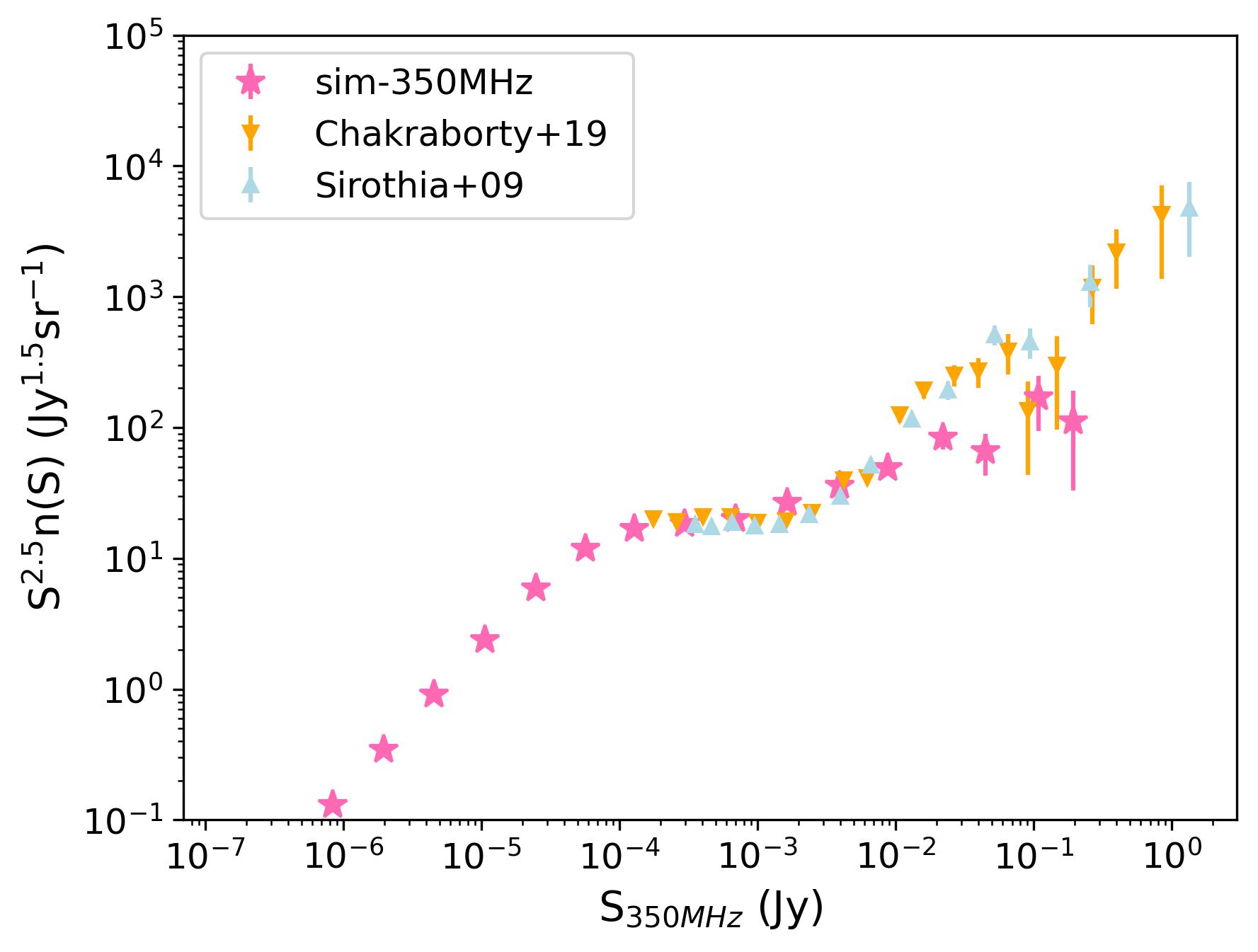}
\end{subfigure}
\begin{subfigure}{0.45\linewidth}
    \includegraphics[width=\linewidth]{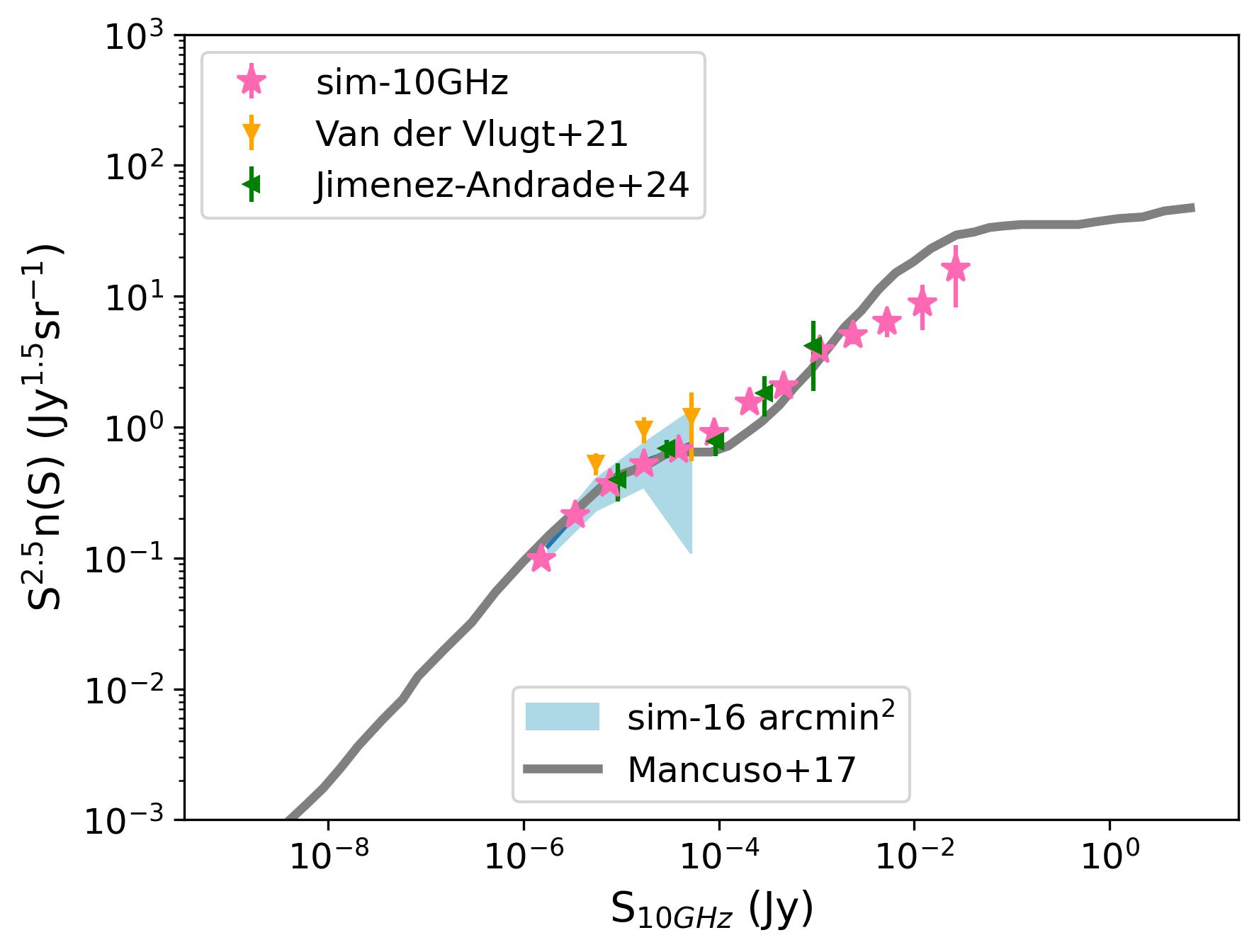}
\end{subfigure}
    
\begin{subfigure}{0.45\linewidth}
    \includegraphics[width=\linewidth]{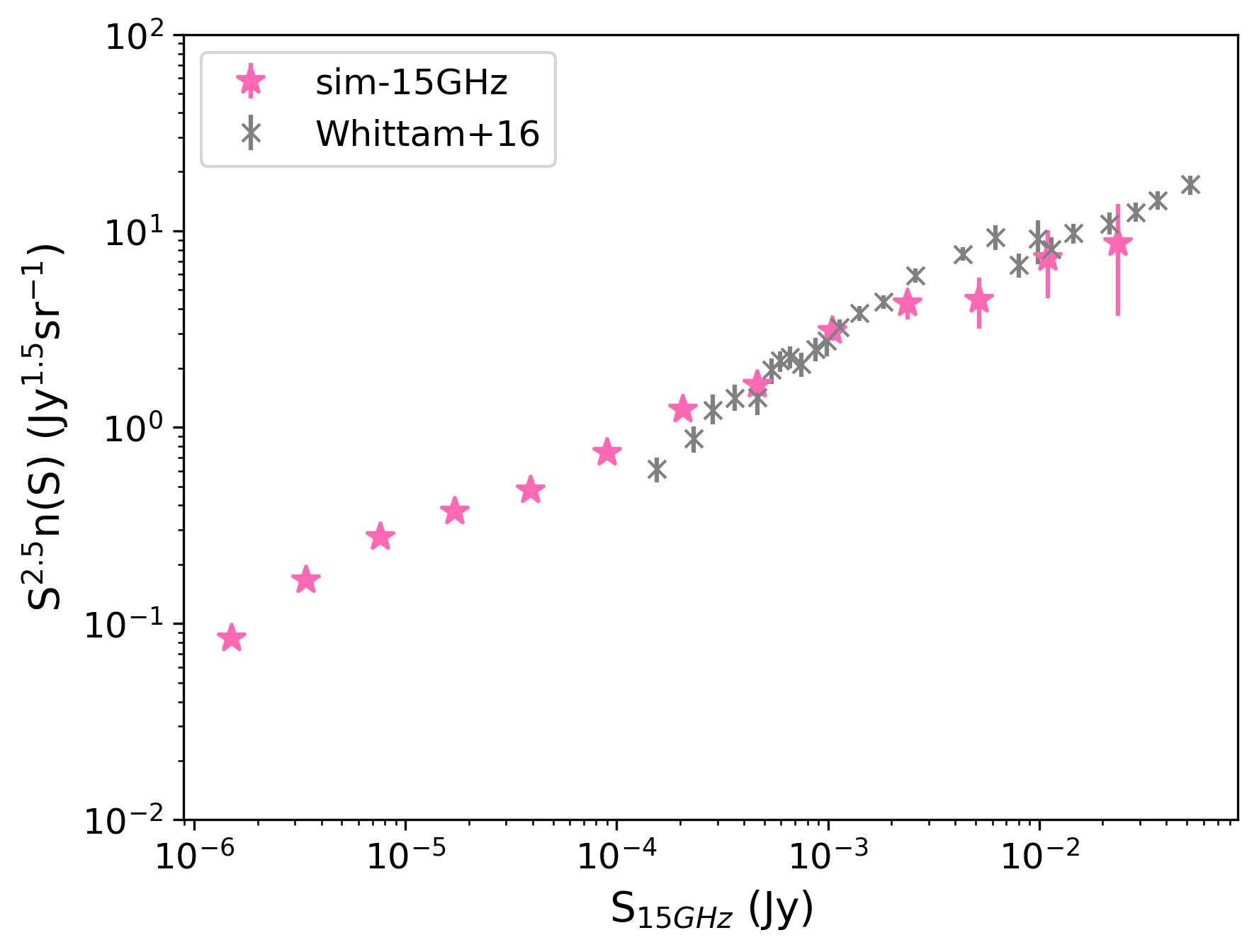}
\end{subfigure}
\caption{Predicted source counts at 350 MHz, 10 GHz and 15 GHz respectively. Theoretical works from \citet{2017ApJ...842...95M} and observations from \citet{2009MNRAS.395..269S,2016MNRAS.457.1496W, 2019MNRAS.490..243C,2021ApJ...907....5V,2024arXiv240613801J} are also included.} 
\label{predict}
\end{figure*}

\subsection{Predictions of source counts at SKA bands }\label{prediction}
In Fig. \ref{predict} we present the predicted source counts at several bands which will be observed by SKA, namely 350 MHz, 10 GHz and 15GHz. As in Section \ref{validation},  we adopt a median spectral slope of $\alpha=-0.5$ for SFGs at 350 MHz and a median spectral slope of $\alpha=-0.5$ for radio AGNs at all three frequencies. 

At 350 MHz, we also include the source counts at 325 MHz and 400 MHz in the ELAIS-N1 field observed through the Giant Metrewave Radio Telescope (GMRT) and the upgraded GMRT (uGMRT) by \citet{2009MNRAS.395..269S} and \citet{2019MNRAS.490..243C} respectively. We correct from 325 MHz and 400 MHz to 350 MHz using $\alpha=-0.5$ for both radio AGNs and SFGs. We find consistent results between our predictions at 350 MHz and both works. 

At 10 and 15 GHz, theoretical work from \citet{2017ApJ...842...95M} and observations from \citet{2016MNRAS.457.1496W}, \citet{2021ApJ...907....5V} and \citet{2024arXiv240613801J} are also included. \citet{2016MNRAS.457.1496W} released deep 15.7 GHz observations through the Arcminute Microkelvin Imager Large Array (AMILA) with best rms noise of 16 $\rm \mu Jy\ beam^{-1}$ in a total area of 0.56 deg$^2$. We do not attempt to correct from 15.7 GHz to 15 GHz due to differences in spectral slopes of the two populations. Based on model-independent approach in \citet{2016ApJ...833..152M,2016ApJ...823..128M}, \citet{2017ApJ...842...95M} derived source counts at 10 GHz. \citet{2021ApJ...907....5V} presented ultra deep 10 GHz observations with a central rms of 0.53 $\rm \mu Jy\ beam^{-1}$ in the COSMOS filed. Since their observations only cover 16 arcmin$^2$, we randomly select an area of the same size in our 4 deg$^2$ simulations and calculate the source counts. We repeat this procedure 100 times to account for cosmic variance. The results are shown as blue shaded region in the right panel of Fig. \ref{predict}. A more recent work by \citet{2024arXiv240613801J} extracted 256 radio sources in the 297 armin$^2$ GOODS-N field. The total 380-hours observation ensured an unparalleled sensitivity of 671 nJy beam$^{-1}$. At these two frequencies, we also find good consistency between our simulations and literature works. Taken all together, our simuations are successful in reproducing observational source counts at various frequencies and can help to design observational strategy, for example, to calculate the observation time needed in order to observe enough faint objects.

\subsection{Demarcation lines between radio AGNs and SFGs}\label{dem}
\begin{figure}[htbp]
\centering
\includegraphics[width=\linewidth]{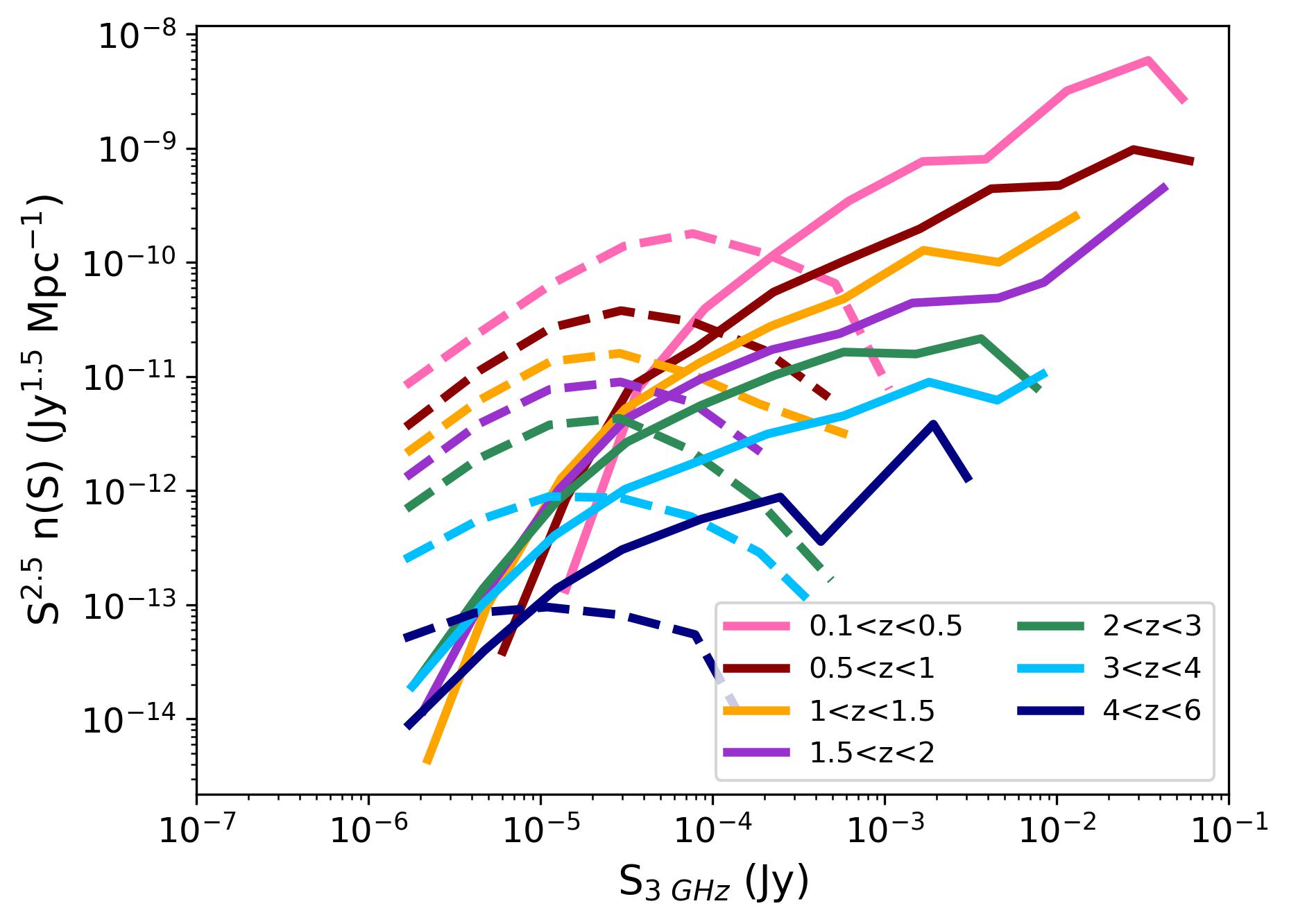}
\caption{Source counts from radio AGNs (solid lines) and SFGs (dashed lines) in different redshift bins at 3 GHz. Other frequencies show similar features.}
\label{demarcation}
\end{figure}

In Fig. \ref{demarcation}, we present normalized source counts at 3 GHz from radio AGNs and SFGs respectively at various redshifts. Demarcation lines exist clearly above which radio AGNs dominate over SFGs. These demarcation lines also evolve with redshift, suggesting higher luminosity threshold of radio AGNs towards higher redshifts. This is reasonable given enhanced SFR in star forming MS galaxies in the early universe, meaning that radio AGNs must possess higher energy to outshine their host galaxies. Other frequencies share similar features. We calculate the demarcation flux density at each band and redshift bin by simply using a powerlaw to fit three nearby data points where number counts from radio AGNs and SFGs intersect. We list them in Table \ref{flux_demar}. These flux densities can be utilized as selection criteria for radio AGNs at different redshifts at different band other than adopting an universal luminosity cut across a wide redshift range.

\begin{table*}[]
    \centering
   \caption{Demarcation flux densities for number counts from radio AGNs and SFGs at different frequency and redshift bin.  }
    \label{flux_demar}
    \begin{tabular}{c c c c c c c c}
    \hline\hline
      frequency & $z<0.5$ & $0.5<z<1$  & $1<z<1.5$ & $1.5<z<2$ & $2<z<3$ & $3<z<4$ & $4<z<6$ \\
      \hline
      150 MHz & 1614.0 & 854.2& 667.4 & 633.0 & 466.1 & 416.8 & 102.8 \\
      350 MHz  & 1105.8 & 495.9 & 468.0 & 400.8 & 319.6 & 250.1 & 95.6 \\
      1.4 GHz & 502.7 & 208.6 & 153.3 & 111.1 & 80.3 & 57.5 & 15.7\\
      3 GHz  & 213.9 & 95.4 & 70.7 & 58.8 & 41.1 & 26.4  & 9.0 \\
      10 GHz & 83.9 & 37.9 & 27.7 & 20.1 & 16.3 & 9.4 & 3.5 \\
      15 GHz  & 68.9 & 31.8 &18.9 & 15.1 & 11.8 & 6.3 & -- \\
      \hline
    \end{tabular}
    \tablefoot{All flux densities are in unit of $\mu$Jy. Due to small number statistics of SFGs at 15 GHz above $z\sim 4$, we remove them as less robust measurements.}
 \end{table*}


\section{Discussion}

\subsection{Deficiencies of our simulation}\label{def}
(1) \textit{The lack of host halo information}
\newline 

Galaxies in simulations are usually accompanied by information of their host halo as dark matter halos are fundamental to their formation and can be designed through $\Lambda$CDM cosmology. For example, in TRECS simulation, \citet{2019MNRAS.482....2B} first generated dark matter halos based on the Planck Millennium simulation \citep{2019MNRAS.483.4922B} and associated them to radio AGNs and SFGs based on their occurrence as a function of dark matter halo mass. However, inherited from EGG, our simulations do not include host halo information and the underlying large-scale structures. Instead, we utilize a special \citet{1978AJ.....83..845S} algorithm to add extra clustering signal of our mock radio AGNs other than simply randomly populating in massive galaxies. Although the dark matter mass can be easily derived from stellar mass to halo mass relation \citep[e.g.,][]{2010ApJ...710..903M,2013ApJ...770...57B,2019MNRAS.488.3143B}, we cannot reproduce clustering signal at larger scale due to limitations of this method, let alone large-scale structures such as filaments. Nonetheless, this lack of large-scale clustering will not affect the main focus of our simulations and is beyond the scope of this letter.\\

(2) \textit{Whether low mass galaxies host low luminosity radio AGNs}
\newline

We assume that radio AGNs residing in host galaxies with stellar mass above $10^{10}\ M_{\odot}$ in Section \ref{assign} which is consistent with literature work that found radio AGNs are mainly hosted by massive galaxies. For example, \cite{2005MNRAS.362...25B} selected 2215 radio-loud AGNs at $0.03<z<0.3$ and found that the fraction of galaxies below $10^{10}\ M_{\odot}$ hosting radio-loud AGNs with $L_{1.4\ GHz}>10^{23}\ \rm W\ Hz^{-1}$ is nearly zero. Nonetheless, we cannot rule out the possibility of low mass galaxies hosting less powerful radio AGNs. We can easily circumvent this issue by simply extrapolating the radio AGN probabilities in Formula \ref{eq1} to lower stellar mass regime. However, it will produce much more radio AGNs since galaxies below $10^{10}\ M_{\odot}$ are way more than galaxies above $10^{10}\ M_{\odot}$: nearly 60\% of EGG galaxies have stellar mass between $10^{8}-10^{10}\ M_{\odot}$, while only $\sim 4$\% of EGG galaxies are above $10^{10}\ M_{\odot}$, under magnitude cut of $F444W=28$ mag. This simple extrapolation will lead to inconsistency of radio AGN number density especially at low redshifts. At higher redshifts this problem is relieved as these low mass galaxies are hardly to be observed given the nominal 3 GHz flux limit of $2.3\ \mu$Jy\ beam$^{-1}$. To keep consistency, we still retain the $10^{10}\ M_{\odot}$ threshold when assign radio AGNs across all redshifts. Whether and how many low mass galaxies host (low luminosity and potentially high luminosity) radio AGNs are an open question for future observational surveys. Hopefully the upcoming deeper SKA observations will help to answer this question. On the other hand, whether the $10^{10}\ M_{\odot}$ threshold is less massive for radio AGNs at higher redshifts is still questionable. However, investigating the boundary stellar mass is beyond the scope of this work. In this work we stick to the $10^{10}\ M_{\odot}$ threshold when assigning radio AGNs. Furthermore, even though these $10^{10}\ M_{\odot}$ galaxies are assigned to host radio AGNs it becomes harder to observe them towards earlier epochs. Only with SKA can these radio-AGN-hosting galaxies with stellar mass around $10^{10}\ M_{\odot}$ be detected as indicated in Fig. \ref{mass_z}. We remain this question to be answered by future work.\\

(3) \textit{The lack of extended  radio sources}
\newline

Our studies are based on radio observations from COSMOS-VLA survey which possesses supreme resolution of 0$\arcsec$.75 \citep{2017A&A...602A...1S}. However, interferometric observations are known to bias against extended radio sources such as sources with elongated radio lobes (FRI and FRII radio loud galaxies). We do not attempt to fix this issue by adding extra FRI and FRII sources due to lack of extensive studies on their fractions as a function of stellar mass and redshift. We caution the use of our simulations in nearby universe. This lack of extended sources may explain partly the lack of extreme source of number counts in Fig. \ref{count}. We hope more extensive data in the future (e.g., SKA observations) would bridge the gap.\\



(4) \textit{mock galaxy and radio AGNs at $z>4$}\label{z4caution}
\newline

The mock galaxy catalog generated by EGG is built upon empirical knowledge on galaxy evolution across cosmic history. So far, our understanding towards high-redshift universe is still largely incomplete due to limited data coverage. To generate mock passive galaxies at $z>4$, EGG code adopts the same parameters of their stellar mass functions at low redshifts, with assumption of a fraction of 15\% given the difficulty to fit individual stellar mass functions with small sample size. When we assign radio AGNs based on AGN probability, we extrapolate redshift-dependent Equations (2) and (5) to $4<z<6$ as \citet{2024arXiv240104924W} only investigated radio AGN probability at $z<4$ due to limited sample size at higher redshifts. For SFGs, the parent sample used in \citet{2021A&A...647A.123D} is restricted to $z<4.5$, so we extrapolate their redshift- and mass- dependent $q_{IR}$ in order to obtain 1.4 GHz radio luminosity for non-AGNs at $z>4.5$. Despite these simple extrapolations, our mock radio catalogs have succeeded in recovering RLFs of both radio AGNs and SFGs at $z>4$, suggesting the universality of these scaling relationships across wide redshift range and the feasibility of our method in reproducing ERB. Upcoming advanced data from new facilities, for example, a more precise constraint of stellar mass functions for passive galaxies at $z>4$ with new JWST observations, would be of great help to the robustness of our mock catalog at high redshifts.

\subsection{Coution on the contribution to ERB from radio AGN and SFGs}\label{contribution}
\begin{figure}[htbp]
\centering
\includegraphics[width=\linewidth]{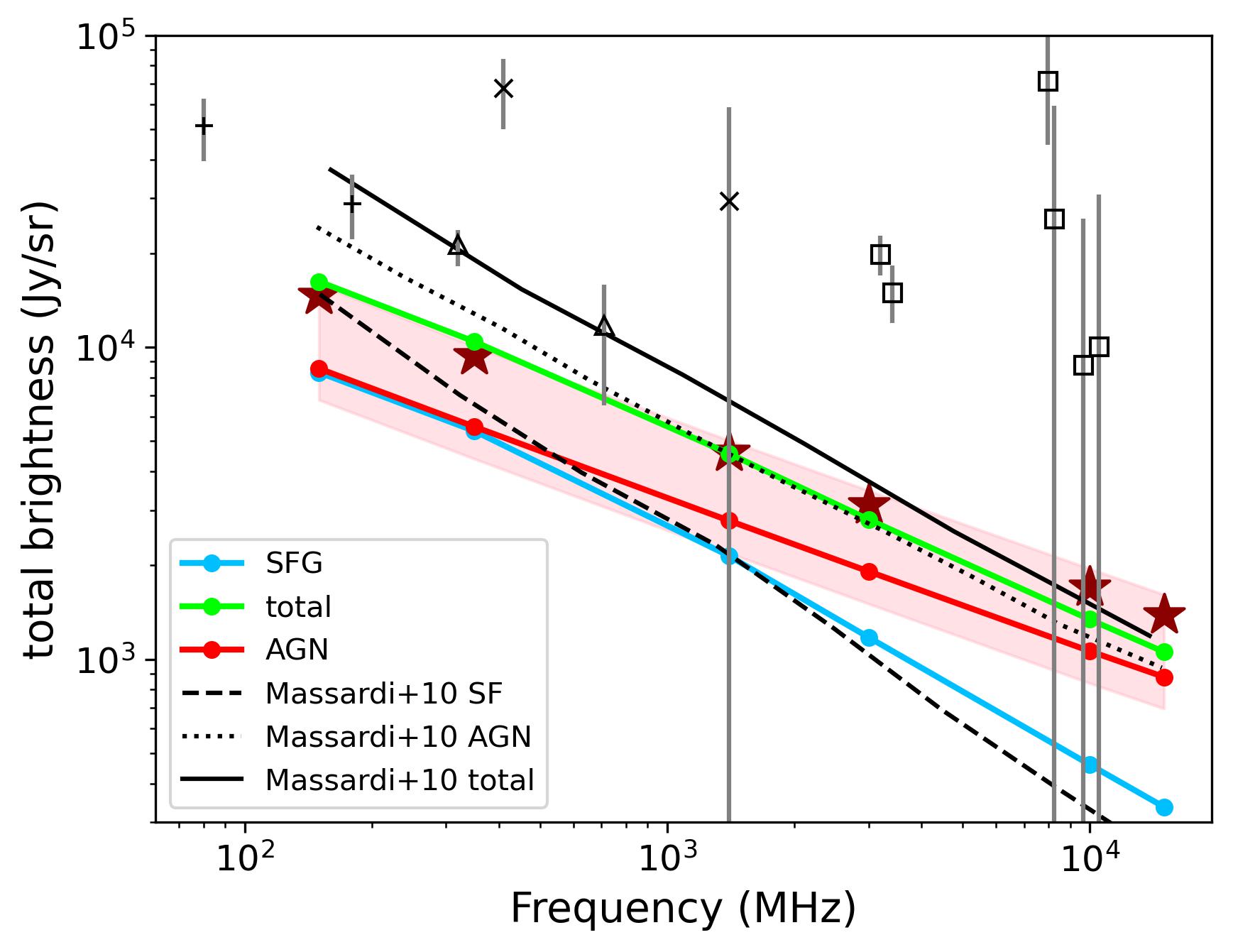}
\caption{Contribution from radio AGNs (red) and SFGs (blue) along with their sum (green) to ERB at various frequencies. We also compare our results with model from \citet{2010MNRAS.404..532M}. Data points are from \citet{1967MNRAS.136..219B, 1970AuJPh..23...45W, 2009ApJ...707..916F} (see Figure 9 in \citealt{2010MNRAS.404..532M}). The red stars denote the contribution from radio AGN plus a single source with $L_{150 MHz}=10^{27}\ \rm W\ Hz^{-1}$ at $z=0.4$. The shaded region represent the 5/95 boundary of radio AGN contribution when running 200 times assigning.}
\label{contri}
\end{figure}

In Fig. \ref{contri} we plot the total flux densities per steradian of both radio AGNs and SFGs along with their sum at various bands. At first glance, the contribution from SFGs and radio AGNs at frequencies higher than 3 GHz are broadly consistent with model predictions in \citet{2010MNRAS.404..532M} and observations, while show significantly lower contribution from radio AGNs towards lower frequencies. Compared to TRECS simulation, we find that this large discrepancy at 150 MHz is mainly attributed to four extremely bright radio AGNs with $L_{150 MHz}>10^{27}\ \rm W\ Hz^{-1}$ at $0.4<z<0.7$ existing in TRECS simulation while lacking in our simulation. Therefore we carry out a test by injecting one radio AGN with $L_{150 MHz}=10^{27}\ \rm W\ Hz^{-1}$ at $z=0.4$ per square degree. The added contributions from radio AGNs are shown as red stars. The significant elevation by merely one source per square degree indicates the caution in comparing radio AGN and SFG contribution especially at low frequencies. Adding one bright AGN will not change their RLF significantly (in fact, radio AGN LF at $0.4<z<0.7$ derived from TRECS simulation is in broad agreement with but slightly higher at the brightest end than observations in \citealt{2022MNRAS.513.3742K}), but their contribution to ERB will be substantially boosted. The driving factors causing one more or fewer bright radio AGNs may be attributed to a biased sample against extended FRII sources (which can reach $L_{150 MHz}=10^{28}\ \rm W\ Hz^{-1}$ at $z \sim 0.4$, e.g., \citealt{2024A&A...683A..23D}) in \citet{2024arXiv240104924W} leading to a smaller fraction of bright radio AGN  calculated via Formula \ref{eq1}. We run a sanity check by assigning radio AGNs 200 times and display the 5/95 percentile distribution of their contribution as shaded pink region. The contribution from radio AGNs is loosely constrained with the majority of them merely differing approximately in one bright source per square degree while the RLF as well as source counts remaining in good agreement with observations. We thereby caution the use of radio AGN contribution to ERB either in theoretical or observational analysis.\\

\section{Summary}\label{summary}

Radio observations play a critical role in help us understanding how galaxies assemble and evolve across the history of our universe. Realistic simulations of extragalactic radio background are important in designing radio surveys and developing radio facilities. Current simulations are mainly built on well-studied RLFs of different populations, despite being successful in recovering ERB, it is difficult to gain insights into other physical information of radio sources and leave the physical origin behind the scene overlooked. We utilize a novel method to associate radio properties with rich physical information of galaxies such stellar mass, SFRs, and multiwavelength flux densities, benefiting the study of galaxy evolution in a synergisitc way.

Based on mock multi-wavelength source catalog generated by EGG code, we assign radio AGNs according to well-defined radio AGN probabilities as a function of radio luminosity, stellar mass and redshift. We assume that radio AGNs only populating in massive galaxies with stellar mass above $10^{10}\ M_{\odot}$ and make use of the star forming main sequence to determine the luminosity threshold above which radio AGN dominates. For SFGs, we take advantage of the tight correlation between IR luminosity and radio luminosity to derive their radio luminosities. Through this method, we manage to link radio flux densities to a plethora of physical information embedded in mock galaxy catalog.

We successfully recover many observables including 1.4 GHz RLFs of both radio AGNs and SFGs, as well as source counts at various frequencies, suggesting that these two populations are important building blocks of ERB. Our successful simulation can help designing observational strategy of radio surveys and evaluating performance of future radio facilities. The link between radio properties and abundant physical properties will benefit studying galaxy evolution in a synergistic way, allowing for exploring multiwavelength counterparts of radio sources. We find a significant fraction of 45--56\% radio sources at $z>4$ that can be detected by SKA, the next generation of radio facilities, will appear invisible in the future deepest optical survey carried out by LSST. This manifests the unparalleled role that deep radio observations play in finding faint dust-obscured galaxies in the early universe. We publicly release our mock sources. The columns are listed in Table \ref{catalog}.
\begin{table*}[]
    \centering
    \caption{Structure of our 4 deg$^2$ mock galaxy catalog. The former 28 columns are generated by EGG \citep[see details in][]{2017A&A...602A..96S}. The later columns are information of our mock radio AGNs and non-AGNs.}
    \label{catalog}
    \begin{tabular}{c c c c}
    \hline\hline
      Column & Name & Unit   & Description \\
      \hline
        1 & id&--& source id\\
        2&ra&degree& source position\\
        3&dec&degree& source position\\
        4&z&--&source redshift\\
        5&mass&M$_{\odot}$&Stellar mass\\
        6&SFR&$\rm M_{\odot}\ yr ^{-1}$ &Star formation rate\\
        7&passive&--&==1 if this source is passive, otherwise active\\
        8&$L_{IR}$&$L_{\odot}$& total IR luminosity\\
        9-28&$f_{bands}$&Jy&flux density at several bands$^{a}$\\
        \hline
        29&AGN&--&==1 if this source is assigned to an AGN\\
        30&qir&--&$q_{IR}$ values for non-AGNs\\
        31&$L_{1.4 GHz}$&W Hz$^{-1}$&rest-frame 1.4 GHz luminosity\\
        32&$\alpha_{\rm low}$&--&spectral slope at 150 and 350 MHz$^{b}$ \\
        33&$\alpha_{\rm high}$&--&spectral slope at 1.4, 3, 10 and 15 GHz$^{b}$ \\
        34-39&$f_{\rm frequency}$&Jy&flux density at multiple frequencies$^{c}$ \\
        40&theta&degree&position angle\\
        41&ellipticity&--&ellipticity of non-AGNs\\
        42&R$_{eff}$&kpc&efftive radius of non-AGNs\\
        \hline
    \end{tabular}
        \tablefoot{ \textit{ a:} \textit{u,g,r,i,z} bands of SDSS, \textit{F606W, F775W, F814W} from HST, F140W, F160W,F444W from JWST, \textit{Ks} band from WIRCAM, 3.6, 4.5, 5.0, 8.0 $\mu$m from IRAC, 24 $\mu$m from MIPS, 250, 350, 500 $\mu$m from \textit{Herschel}. \textit{b:} The spectral slope of AGNs at low and high frequencies are the same. \textit{c:} We provide flux densities at 150 MHz, 350 MHz, 1.4 GHz, 3 GHz, 10 GHz and 15 GHz. Flux densities at other frequencies can be easily derived from $L_{1.4 GHz}$.}
\end{table*}

\section*{Acknowledgement}
We thank the anonymous referee for a through and constructive report. This work was supported by National Natural Science Foundation of China (Project No. 12173017 and Key Project No. 12141301), National Key R\&D Program of China (2023YFA1605600) and the China Manned Space Project (No. CMS-CSST-2021-A07). F.Y.G acknowledges supports by Jiangsu Provincial Outstanding Postdoctoral Program ([2023]209) and Yuxiu Young Scholars Program in Nanjing University. Y.J.W. acknowledges supports by National Natural Science Foundation of China (Project No. 12403019) and the National Science Foundation of Jiangsu Province (BK20241188).

\bibliographystyle{aa}
\bibliography{ref.bib}

\end{document}